\pdfoutput=1
\documentclass[11pt,a4paper]{article}
\usepackage{jheppub}
\usepackage{overpic}
\usepackage{marvosym}
\usepackage{hyperref}
\usepackage[bottom]{footmisc}

\newcommand{\pT}{\ensuremath{p_\mathrm{T}}}
\newcommand{\pythia}{{\sc{pythia}}}
\newcommand{\phojet}{{\textsc{phojet}}}
\newcommand{\herwig}{{\textsc{herwig}}}

\usepackage{preprintcover} 
\PreprintCoverPaperTitle{Measurement of inclusive two-particle angular correlations \\ in pp collisions with the ATLAS detector at the LHC} 
\PreprintIdNumber{CERN-PH-EP-2012-011}  
\PreprintCoverAbstract{We present a measurement of two-particle angular correlations in proton-proton collisions at $\sqrt{s} =$ 900 GeV and 7 TeV. The collision events were collected during 2009 and 2010 with the ATLAS detector at the Large Hadron Collider using a single-arm minimum bias trigger. Correlations are measured for charged particles produced in the kinematic range of transverse momentum $\pT >$ 100~MeV and pseudorapidity $|\eta| <$~2.5. A complex structure in pseudorapidity and azimuth is observed at both collision energies. Results are compared to \pythia~8 and \herwig++ as well as to the AMBT2B, DW and Perugia 2011 tunes of \pythia~6.  The data are not satisfactorily described by any of these models.}  
\PreprintJournalName{JHEP}

\title{Measurement of inclusive two-particle angular correlations in pp collisions with the ATLAS detector at the LHC}

\author{The ATLAS collaboration}

\abstract{
We present a measurement of two-particle angular correlations in proton-proton collisions at $\sqrt{s} =$ 900 GeV and 7 TeV. The collision events were collected during 2009 and 2010 with the ATLAS detector at the Large Hadron Collider using a single-arm minimum bias trigger. Correlations are measured for charged particles produced in the kinematic range of transverse momentum $\pT >$ 100~MeV and pseudorapidity $|\eta| <$~2.5. A complex structure in pseudorapidity and azimuth is observed at both collision energies. Results are compared to \pythia~8 and \herwig++ as well as to the AMBT2B, DW and Perugia 2011 tunes of \pythia~6.  The data are not satisfactorily described by any of these models.
}

\keywords{Hadron-Hadron Scattering}

\begin{document}
\maketitle
\flushbottom

\section{Introduction}
\label{sec:introduction}

Several phenomenological models have been formulated in order to describe the dynamics of multi-particle production in high-energy hadron collisions.  Examples include  the string \cite{Andersson:1987} and cluster \cite{Morel:1974} models,  but all have proven to be incomplete since they provide only limited explanations for the observed production of soft particles. The study of correlations between final-state particles is a powerful method for investigating the underlying mechanisms of particle production, as demonstrated by previous experiments \cite{Eggert:1975,UA5:1988,PHOBOS:2007}.  Phenomenological models of soft particle production are encapsulated in a variety of different Monte Carlo (MC) generators. Particle correlation measurements provide useful input for discriminating between such models and for tuning their parameters in order to better describe non-perturbative physics. Model features that are sensitive to measurements of correlations include fragmentation, resonance decays, multi-parton interactions, and the emergence of jets. The measurement of correlated particle production in proton-proton collisions is also very useful for studies of heavy ion collisions; any deviation in the latter from the behaviour seen in the former can be attributed to the fact that partons experience additional interactions, and other collective effects, in the hot and dense medium present in heavy ion collisions.

This study focuses on the two-particle angular correlations in pseudorapidity, $\eta$, and azimuthal angle, $\phi$, using proton-proton collision data collected by the ATLAS experiment\footnote[1]{ATLAS uses a right-handed coordinate system with its origin at the nominal interaction point (IP) in the centre of the detector and the $z$-axis along the beam pipe. The $x$-axis points from the IP to the centre of the LHC ring, and the $y$-axis points upwards. Cylindrical coordinates $(r,\phi)$ are used in the transverse plane, $\phi$ being the azimuthal angle around the beam pipe. The pseudorapidity is defined in terms of the polar angle $\theta$ as $\eta=-\ln\tan(\theta/2)$.} at the Large Hadron Collider (LHC).  A measurement of the inclusive two-particle angular correlation function is performed for charged particles produced in the kinematic region defined by $\pT>100$\ MeV~\footnote[2]{The quantity $\pT$ corresponds to the component of the charged-particle momentum transverse to the beam direction.} and $|\eta|<2.5$ in proton-proton collisions at centre-of-mass collision energies of $\sqrt{s} =$ 900 GeV and 7 TeV. For the 7 TeV centre-of-mass energy, an additional region of phase space is considered for which the charged-particle multiplicity per event is at least twenty.  

The paper is organised as follows: the two-particle correlation definition used here is described in section \ref{sec:definition}, while a brief description of the ATLAS detector is presented in section \ref{sec:detector}. Section \ref{sec:montecarlo} summarises the different Monte Carlo models used in this study and the data samples and event selection are described in section \ref{sec:selection}. In section \ref{sec:correction} the data-driven procedure for unfolding the effect of the detection apparatus from the physics is presented. The sources of systematic uncertainty affecting the analysis are discussed in section \ref{sec:systematics}. Finally, the corrected two-particle angular correlation functions and a comparison to a variety of different Monte Carlo predictions are shown in section \ref{sec:results}.

\section{Definition of the two-particle correlation}
\label{sec:definition}

The correlation function is defined using two different distributions that are dependent on $\eta$ and $\phi$. The foreground distribution, $F(n_{ch}, \Delta\eta,\Delta\phi)$, describes the angular separation in azimuth ($\Delta\phi$) and pseudorapidity ($\Delta\eta$) between pairs of particles emitted in the same event of charged multiplicity $n_{ch}$.  $F(n_{ch}, \Delta\eta, \Delta\phi)$ is defined for all events with $n_{ch}$ greater than one and  contains correlations due to both the underlying physics processes and detector effects such as geometrical acceptance. The expression for $F(n_{ch}, \Delta\eta, \Delta\phi)$ is given by 

\begin{equation}
F\left(n_{ch}, \Delta\eta, \Delta\phi\right)=\left<\frac{2}{n_{ch}\left(n_{ch}-1\right)}\sum_{i}\sum_{j\ne i}\delta(\eta_{i}-\eta_{j}-\Delta\eta)\delta(\phi_{i}-\phi_{j}-\Delta\phi)\right>,\label{eqn:F}
\end{equation}

\noindent where the summation is over all charged particles in a single event and the average is taken over an ensemble of many events, all having the same charged-particle multiplicity of $n_{ch}$. The delta functions, $\delta(\eta_{i}-\eta_{j}-\Delta\eta)$ and $\delta(\phi_{i}-\phi_{j}-\Delta\phi)$, select particle pairs with the $\Delta\eta$ and $\Delta\phi$ separation that is appropriate for $F(n_{ch}, \Delta\eta, \Delta\phi)$.  The total number of charged-particle pairs in an event is $n_{ch}\left(n_{ch} - 1\right)/2 $, therefore if the factor of $2/n_{ch}\left(n_{ch} - 1\right)$ were not included then a particle in a low multiplicity event would carry a lower weight than an otherwise identical particle in a higher multiplicity event because there are fewer possible pair combinations.  

Since the average charged-particle distribution, $\mathrm{d}n/\mathrm{d}\eta$, is approximately flat in $\eta$ within the acceptance of the measurement,  phase-space alone dictates that the $\Delta\eta$ distribution is peaked at $\Delta\eta=0$ and falls approximately linearly to a maximum possible value of $\Delta\eta=\pm2\eta_{max}$. This motivates the second distribution used in constructing the correlation function, the background correlation $B$, the $\Delta\eta$ dependence of which at a fixed charged-particle multiplicity is given by

\begin{equation}
\begin{split}
B\left(n_{ch}, \Delta\eta\right) =& \int_{-\eta_{max}}^{+\eta_{max}}\int_{-\eta_{max}}^{+\eta_{max}}\mathrm{d}\eta_{1}\mathrm{d}\eta_{2}\delta\left(\eta_{1} - \eta_{2} - \Delta\eta\right)\\ & \left.\frac{1}{n_{ch}}\frac{\mathrm{d}n\left(n_{ch}, \eta\right)}{\mathrm{d}\eta}\right|_{\eta=\eta_{1}}\left.\frac{1}{n_{ch}}\frac{\mathrm{d}n\left(n_{ch},\eta\right)}{\mathrm{d}\eta}\right|_{\eta=\eta_{2}},\label{eqn:B}
\end{split}
\end{equation}

\noindent where $\mathrm{d}n\left(n_{ch}, \eta\right)/\mathrm{d}\eta$ is the $\eta$ distribution of charged particles in events of fixed charged-particle multiplicity $n_{ch}$ and the factors of $1/n_{ch}$ ensure that the integral of $B$ over all $\Delta\eta$ is unity. The function $\delta(\eta_{1}-\eta_{2} -\Delta\eta)$ is a Dirac delta function.  The $\Delta\phi$ dependence of $B$ is defined analogously to, and is flatter than, the $\Delta\eta$ dependence.

The correlation function $C(n_{ch},\Delta\eta,\Delta\phi)$ can be constructed using the foreground and background distributions, and is 

\begin{equation}
C(n_{ch},\Delta\eta,\Delta\phi) = F(n_{ch},\Delta\eta,\Delta\phi) - B(n_{ch},\Delta\eta,\Delta\phi).\label{eqn:C}
\end{equation}

Previous analyses \cite{Eggert:1975} have found that $(n_{ch}-1) \times C(n_{ch},\Delta\eta,\Delta\phi)$ is approximately independent of $n_{ch}$.  By multiplying equation (\ref{eqn:C}) by a factor of $\left(n_{ch} - 1\right)$ and averaging the resulting quantity over all charged particle multiplicities greater than one, we obtain the multiplicity-independent result, $\overline{C}$, given by

\begin{align}
\overline{C}\left(\Delta\eta, \Delta\phi\right) &=  \sum_{n_{ch}}P\left(n_{ch}\right)\left(n_{ch} - 1\right)\left\{F\left(n_{ch}, \Delta\eta, \Delta\phi\right)-B\left(n_{ch}, \Delta\eta, \Delta\phi\right)\right\} \notag\\
 &=\left<\left(n_{ch}-1\right)F\left(n_{ch}, \Delta\eta, \Delta\phi\right)\right>_{ch} - \left<\left(n_{ch}-1\right)B\left(n_{ch}, \Delta\eta, \Delta\phi\right)\right>_{ch}, \label{eqn:Cbar}
\end{align}
where $P\left(n_{ch}\right)$ is the probability density function for $n_{ch}$ and the notation $\left<...\right>_{ch}$ indicates that a weighted average over contributions from all charged-particle multiplicities should be made.

Dividing equation (\ref{eqn:Cbar}) by the multiplicity-averaged background distribution then gives the multiplicity-independent two-particle correlation function

\begin{equation}
R\left(\Delta\eta, \Delta\phi\right) = \frac{\left<\left(n_{ch}-1\right)F\left(n_{ch}, \Delta\eta, \Delta\phi\right)\right>_{ch}}{\left<B\left(n_{ch}, \Delta\eta, \Delta\phi\right)\right>_{ch}} - \frac{\left<\left(n_{ch} - 1\right)B\left(n_{ch}, \Delta\eta, \Delta\phi\right)\right>_{ch}}{\left<B\left(n_{ch}, \Delta\eta, \Delta\phi\right)\right>_{ch}}.\label{eqn:R}
\end{equation}

Since the $\mathrm{d}n/\mathrm{d}\eta$ distribution, and hence the background, is not \emph{strongly} dependent on the charged particle multiplicity, the factor of $B$ approximately cancels from the second term on the right of equation\ (\ref{eqn:R}) with two caveats. First, the background \emph{does} have some multiplicity dependence at low multiplicity due to the contribution from double diffraction, which favours higher $\Delta\eta$ values compared to the non-diffractive contribution. Second, the effect of any track reconstruction inefficiency is larger in low multiplicity events compared to high multiplicity events because a small number of mis-reconstructed tracks has a proportionally larger significance in the former. As an example, consider a low multiplicity event with a small number of correlated track clusters; the loss of a small fraction of tracks could easily remove completely one of the track clusters, thus changing the overall correlation.  The loss of the same fraction of tracks in a higher multiplicity event is less likely to remove such a feature; it will instead simply reduce the overall number of track pairs. Calculating a multiplicity independent background distribution, $B\left(\Delta\eta, \Delta\phi\right)$,  using $\left<\mathrm{d}n/\mathrm{d}\eta\right>_{ch}$, which is already averaged over all multiplicities, therefore has the advantage of diluting the effect of the experimentally more troublesome lower multiplicity events. Such an approach slightly reduces the sensitivity of the observable to diffraction. However, diffraction is not the motivation for this measurement. The final expression used for the inclusive two-particle correlation function is then given by 

\begin{equation}
R\left(\Delta\eta, \Delta\phi\right) = \frac{\left<\left(n_{ch}-1\right)F\left(n_{ch}, \Delta\eta, \Delta\phi\right)\right>_{ch}}{B\left(\Delta\eta, \Delta\phi\right)} - \left<n_{ch} - 1\right>_{ch}.\label{eqn:RFinal}
\end{equation}

In practice, the expression $\left<\left(n_{ch}-1\right)F\left(n_{ch}, \Delta\eta, \Delta\phi\right)\right>_{ch}$ is constructed by taking each pair of particles within a single event, calculating their absolute $\eta$ and $\phi$ separations and filling one quadrant of a two-dimensional distribution at those values using a weight of $2/n_{ch}$. The other three quadrants are filled by reflection, making the distribution symmetric around (0,0). This distribution is normalised by dividing each bin by the number of events entering the distribution. The background is determined by taking random pairs of events and, for each particle in one event, the $|\Delta\eta|$ and $|\Delta\phi|$ values with each particle in the other event are calculated and used to fill another two-dimensional distribution, in the same way as done to the foreground distribution, which is then normalised to unit integral. 

Projections of the two-dimensional correlation function along both $\Delta\eta$ and $\Delta\phi$ help reveal more details of the structure of the correlations and permit easier comparisons with different models. These projections are calculated by first integrating separately the foreground and the background distributions before taking the ratio between the two, and normalising with the average track multiplicity.

\section{The ATLAS detector}
\label{sec:detector}

The ATLAS detector \cite{ATLAS:2008} is one of the two general-purpose detectors at the LHC \cite{LHC:2008}. Collision events are reconstructed by layers of tracking, calorimeter and muon systems covering almost the whole solid angle around the collision point. For the analysis presented in this paper, the tracking detectors and trigger systems are the most relevant.

At the centre of ATLAS lies the inner detector (ID), which is responsible for reconstructing the trajectories and measuring the transverse momenta of charged particles and reconstructing the vertices from which they originate. The ID has complete azimuthal coverage and consists of three sub-systems, which cover the pseudorapidity range of $|\eta| < 2.5$. The three ID sub-systems are: a silicon pixel detector, a semiconductor tracking detector (SCT) and a transition radiation tracker (TRT). They are immersed in a 2 T axial magnetic field and cover a radial distance from the nominal interaction point in the barrel of 50.5-150 mm, 299-560 mm and 563-1066 mm, respectively. The ID sensors provide a position resolution of 10, 17 and 130 $\mu$m for the $r-\phi$ coordinate and of 115 and 580 $\mu$m for the second coordinate in the case of the pixel and SCT detectors. The angular resolution of the tracks reconstructed from the pixel, SCT and TRT hits is between 40 $\mu$rad (for $\eta\simeq0.25$) and 50 $\mu$rad (for $\eta\simeq1.75$) in azimuth and between 5 and 10 in $\cot\left(\theta\right)$.

There are three levels in the ATLAS trigger system: Level-1, Level-2 and the Event Filter. The first level uses limited information to perform an initial selection of potentially interesting events, with a fixed latency of 2.5~$\mu$s. The subsequent levels use full granularity measurements to further reduce the selected events to a level acceptable for offline reconstruction. For this measurement, only the Level-1 trigger signals from the beam pickup timing devices (BPTX) and minimum bias trigger scintillators (MBTS) are used to select candidate events. The BPTX electrostatic button pick-up detectors are attached to the beam-pipe and located at $\pm175$~m from the collision point. The MBTS consists of plastic scintillators on the inner edge of the calorimeter end-caps, which lie on both sides of the detector, each at a distance of 3.56~m from the interaction point. Segmented into eight sectors in azimuth, they have a pseudorapidity coverage of $2.09 < |\eta| < 2.82$ and $2.82 < |\eta| < 3.84$. A coincidence of the BPTX signals between the two sides of ATLAS (i.e. colliding bunches) and at least one hit above threshold from either side of the MBTS (referred to as single-arm trigger) are required.

\section{Monte Carlo models}
\label{sec:montecarlo}

In order to compare the present measurement to different soft QCD models, three MC event generators are considered: \pythia~6.4 (versions 6.421 and 6.425) \cite{Pythia6:2006}, \pythia~8.150 \cite{Pythia8:2008} and \herwig++ version~2.5.1 \cite{Herwig:2008}.

The \pythia\ approach to soft interactions uses phenomenological adaptations of the lowest-order perturbative QCD to describe low-$\pT$ scattering processes in order to extend its physics description to the non-perturbative regime.\ \herwig++, on the other hand, extends its eikonal concept of independent multi-partonic scatters down into the infrared region \cite{HSoft:2009}.  Other differences between \herwig++ and \pythia\ include the hadronisation model and the ordering of the parton shower. \pythia\ uses a hadronisation model based upon colour strings and a \pT\ or virtuality-ordered shower, while \herwig++ implements a cluster hadronisation scheme and a shower that is ordered in emission angle.

Each Monte Carlo generator provides a set of parameters whose values can be altered to produce different MC tunes.  Several tunes of \pythia\ version 6.4 are compared to data: AMBT2B \cite{AMBT:2011}, Perugia 2011 \cite{Per11:2011} and DW \cite{DW:2002}. Version 6.425 of \pythia\ was used in all cases apart from tune DW at 900~GeV, for which version 6.421 was used\footnote[3]{The small differences between the minor \pythia\ versions do not affect the non-perturbative physics relevant for this measurement}.  Tune 4C of \pythia\ 8.150  \cite{P84C:2011} as well as the UE7-2 underlying-event tune at $\sqrt{s} =$ 7 TeV and the MU900-2 minimum bias tune at $\sqrt{s} =$ 900 GeV \cite{HTunes:2011} of \herwig++ \cite{H25:2011} are also used for comparison. 

AMBT2B is a tune of \pythia\ 6 that was produced by the ATLAS Collaboration as a result of optimising earlier tunes to best describe ATLAS results.  In particular, parameters controlling the initial state radiation (ISR) cut-off and evolution, plus further parameters that adjust the model for multi-parton interactions (MPI) were modified to reproduce data on jet shapes and minimum bias events from ATLAS. The tune AMBT2B with the CTEQ6L1 parton density functions (PDFs)  \cite{CTEQ6:2002} is used here, and provides one of the best available descriptions of existing minimum bias data.

The Perugia series of tunes predates the availability of LHC data; however, the most recent Perugia 2011 tune has been adjusted to take account of measurements made at the LHC.  Differences between the Perugia 2011 and AMBT2B tunes include the ISR evolution and the parameters controlling the reconnection of outgoing string fragments to the beam remnants.  Perugia 2011 is shown here with the CTEQ 5L PDFs.

Both AMBT2B and Perugia 2011 use a $\pT$-ordered parton shower and a model in which multi-parton interactions are interleaved with ISR. DW, on the other hand, is a tune that uses a virtuality-ordered parton shower and a MPI model that is not interleaved with the initial state shower. It is optimised to describe underlying event and Drell-Yan data from the Tevatron. 

\pythia\ 8 adds to the MPI model of \pythia\ 6 by interleaving not only the ISR, but also the final state radiation (FSR) with the MPI.  \pythia\ 8 includes an updated model of diffraction that includes harder colour singlet exchange;  compared to \pythia\ 6, this allows for harder $\pT$ and particle multiplicity distributions from the single- and double-diffractive contributions.  Like the version of AMBT2B shown here, tune 4C of \pythia\ 8 uses the CTEQ6L1 PDFs.   While \pythia\ 8 has not yet been tuned to data as extensively as \pythia\ 6, tune 4C was the result of comparisons to early LHC data after \pythia\ 8 had already been tuned to give reasonable agreement with Tevatron minimum bias data.  The result is that tune 4C has a reduced diffractive contribution and more realistic average \pT\ per particle compared to earlier tunes of \pythia\ 8.  

\herwig++ is used here with the MRST2007LO* PDFs \cite{MRTS:2008}. The current version of \herwig++ does not have a model for diffraction, so it has been tuned to diffraction-suppressed LHC data. \herwig++ has a model for hadronisation that is based on cluster decay, rather than the string fragmentation used by \pythia.  It is therefore interesting to test \herwig++ against the present measurement in order to compare the two hadronisation models and to ascertain the importance of a diffractive model in describing two-particle correlations.

In addition to these models, the older Perugia 0 \cite{Per0:2009} and ATLAS MC09 \cite{MC09:2010} \pythia\ 6 tunes, the Tune 1 of \pythia\ 8.130 and the \phojet\ \cite{Phojet:1995} generator are used for systematic studies.

Samples containing non-diffractive and diffractive components, mixed according to the generator cross sections, are used when appropriate for those models that provide a separate diffractive contribution. Particles with a mean decay length ($c\tau$) greater than 10~mm are not decayed by the generator. As a consequence, secondary particles (such as the decay products of neutral kaons) are excluded from the generator-level distributions. This setting also defines the hadronic final state to which the measured distributions are corrected. The generated events are processed through the GEANT~4 \cite{Geant4:2002} based ATLAS detector simulation program \cite{ATLASsim:2010}.  In addition to simulating the detector, GEANT~4 also handles the decays of particles whose mean decay length is longer than 10~mm. Having been fully simulated, MC events are then reconstructed and analysed in the same manner as the data.

\section{Data samples and event selection}
\label{sec:selection}

This study uses the same selection criteria as the charged-particle multiplicity analysis carried out at $\sqrt{s} =$ 900~GeV and 7~TeV \cite{MB2:2011}. In total, approximately 7~$\mu$b$^{-1}$  of 900~GeV data and 190~$\mu$b$^{-1}$ of 7~TeV data are analysed. The events were collected during 2009 and 2010 using low instantaneous luminosity proton beams such that the mean number of interactions per bunch crossing was around 0.01. Only events in which the inner detector was fully operational and the solenoidal magnet was on are used. 

Tracks are selected for analysis if they have:
\begin{itemize}
		\item $\pT >$ 100 MeV,
		\item $|\eta| <$ 2.5,
		\item at least 1 hit in the innermost layer of the pixel detector,
		\item at least 1 hit in any of the layers of the pixel detector,
		\item at least 2, 4 or 6 hits in the SCT for tracks with $\pT > $~100,~200 and 300~MeV, respectively,
		\item a transverse impact parameter with respect to the primary vertex, $d_{0}^{\mathrm{PV}}$,  of $|d_{0}^{\mathrm{PV}}| <$ 1.5 mm,
		\item a longitudinal impact parameter with respect to the primary vertex, $z_{0}^{\mathrm{PV}}$, of $|z_{0}^{\mathrm{PV}}\sin\theta| <$ 1.5 mm (where $\theta$ is the angle between the track and the $z$-axis),
		\item a track-fit $\chi^{2}$ probability of at least 0.01 for those tracks with  $\pT >$ 10 GeV (to remove tracks with mis-measured $\pT$ due to misalignment or interactions within the detector material).
	\end{itemize} 

To reduce the contribution from background events (such as beam-halo, beam-gas interactions and cosmic rays) and non-primary tracks, events are required:
\begin{itemize}
	\item to have been triggered by the single-arm trigger, 
	\item to contain at least one primary vertex.  Primary vertices are required to be associated with at least two tracks with $\pT >$ 100 MeV with a transverse impact parameter with respect to the beam spot\footnote[4]{The beam spot is defined as the centre of the luminous region in which the proton beams interact.} of  $\left|d_{0}^{\mathrm{BS}}\right| <$ 4 mm,
	\item not to have a second interaction vertex associated to more than four tracks (to remove events with more than one proton-proton interaction in a single bunch crossing),
	\item to contain a minimum number of tracks, depending on the phase-space region, passing the track selection criteria previously described.
\end{itemize}

For data taken at $\sqrt{s} =$ 7 TeV, two different phase-space regions with varying contributions from diffractive events are explored: $n_{ch} \geq 2$ and $n_{ch} \geq 20$. For data taken at $\sqrt{s} =$ 900 GeV, only the particle multiplicity requirement of $n_{ch} \geq$ 2 is used. The total numbers of selected events and tracks are shown in table \ref{tab:Nevents_data}.

\begin{table}[t]
\begin{tabular}{|l|c|c|c|}
\hline \hline
Phase-Space Region						& $\sqrt{s}$ [TeV]	& Selected Events	& Selected Tracks	\\
\hline
$n_{ch}\geq2$, $\pT >$ 100 MeV, $|\eta| <$ 2.5	& 0.9			& 357,523			& 4,532,663		\\
\cline{2-4}
		         							& 7				& 10,066,072		& 209,809,430		\\
\hline
$n_{ch}\geq20$, $\pT >$ 100 MeV, $|\eta| <$ 2.5	& 7				& 4,029,565		& 153,553,766       	\\
\hline \hline
\end{tabular}
\centering
\caption{Total number of selected events and tracks in 900 GeV and 7 TeV data.}
\label{tab:Nevents_data}
\end{table}

\section{Correction procedure}
\label{sec:correction}

In order to disentangle the effects caused by the apparatus and the reconstruction algorithms from the true physics processes, the following corrections are applied to the reconstructed data.

First, the foreground and multiplicity distributions are weighted with the inverse of the trigger and vertex reconstruction efficiencies. These efficiencies, taken directly from ref.~\cite{MB2:2011}, are calculated before a requirement on the vertex is made. The efficiencies were determined in ref.~\cite{MB2:2011} from data by using a trigger independent of the Level 1 MBTS (for the trigger efficiency) and by relaxing the requirement on the presence of a reconstructed vertex (for the vertex efficiency). These efficiencies depend on the track multiplicity, where the total number of selected tracks is obtained by applying all of the selection criteria  described in section \ref{sec:selection} apart from  the $d_{0}^{\mathrm{PV}}$ and $z_{0}^{\mathrm{PV}}\sin\theta$ requirements.  A requirement on the transverse impact parameter with respect to the beam spot position, $d_{0}^{\mathrm{BS}}$, is made instead such that $|d_{0}^{\mathrm{BS}}| <$ 1.8 mm.  The trigger efficiency is determined using a test trigger that selects events from random proton bunch crossings and uses information from the pixel and SCT detectors at Level 2.

\begin{figure}[t]
\begin{center}
\begin{overpic}[height=53.5mm]{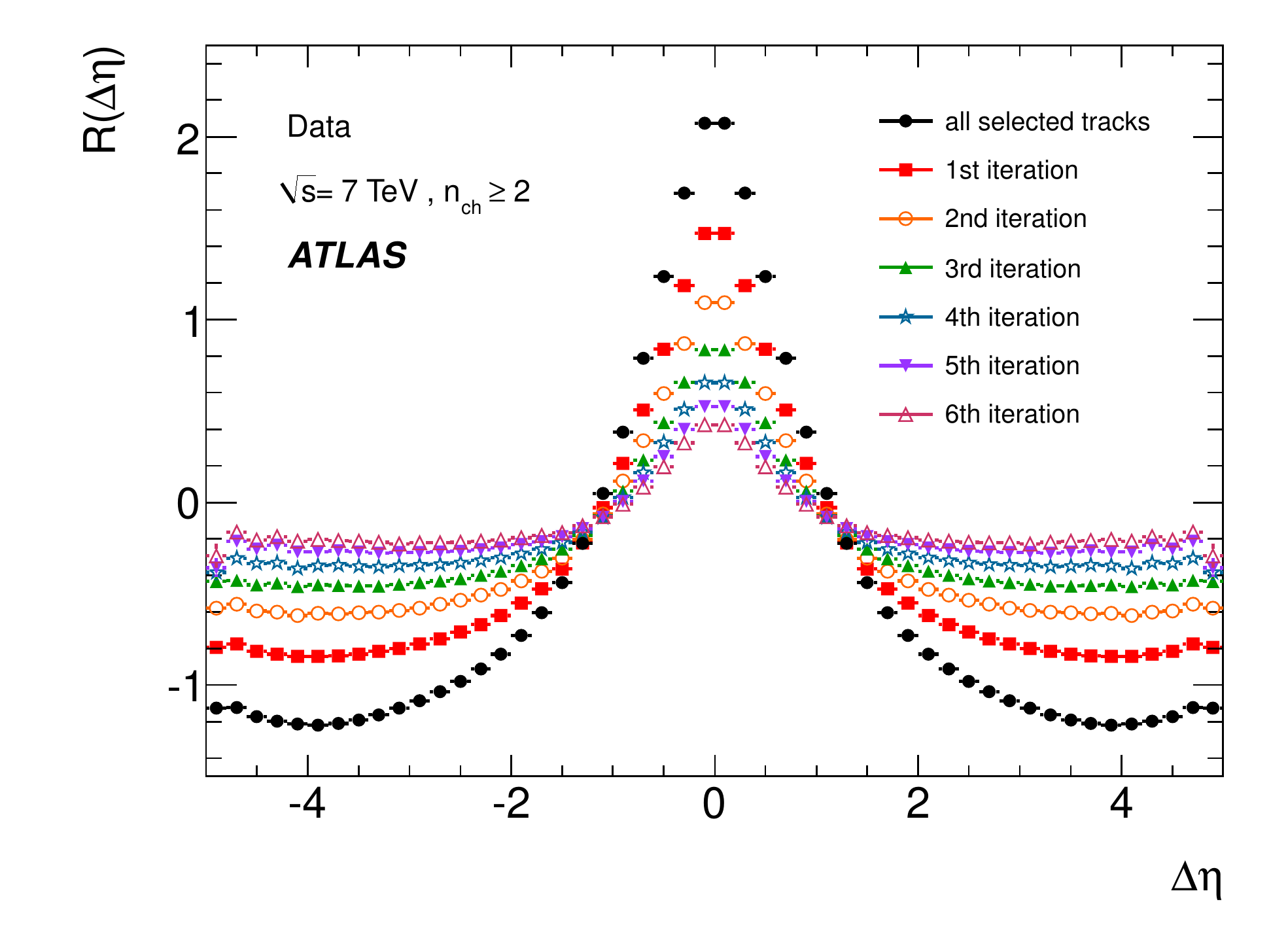}
\put(53,-3){\small(a)}
\end{overpic}
\begin{overpic}[height=53.5mm]{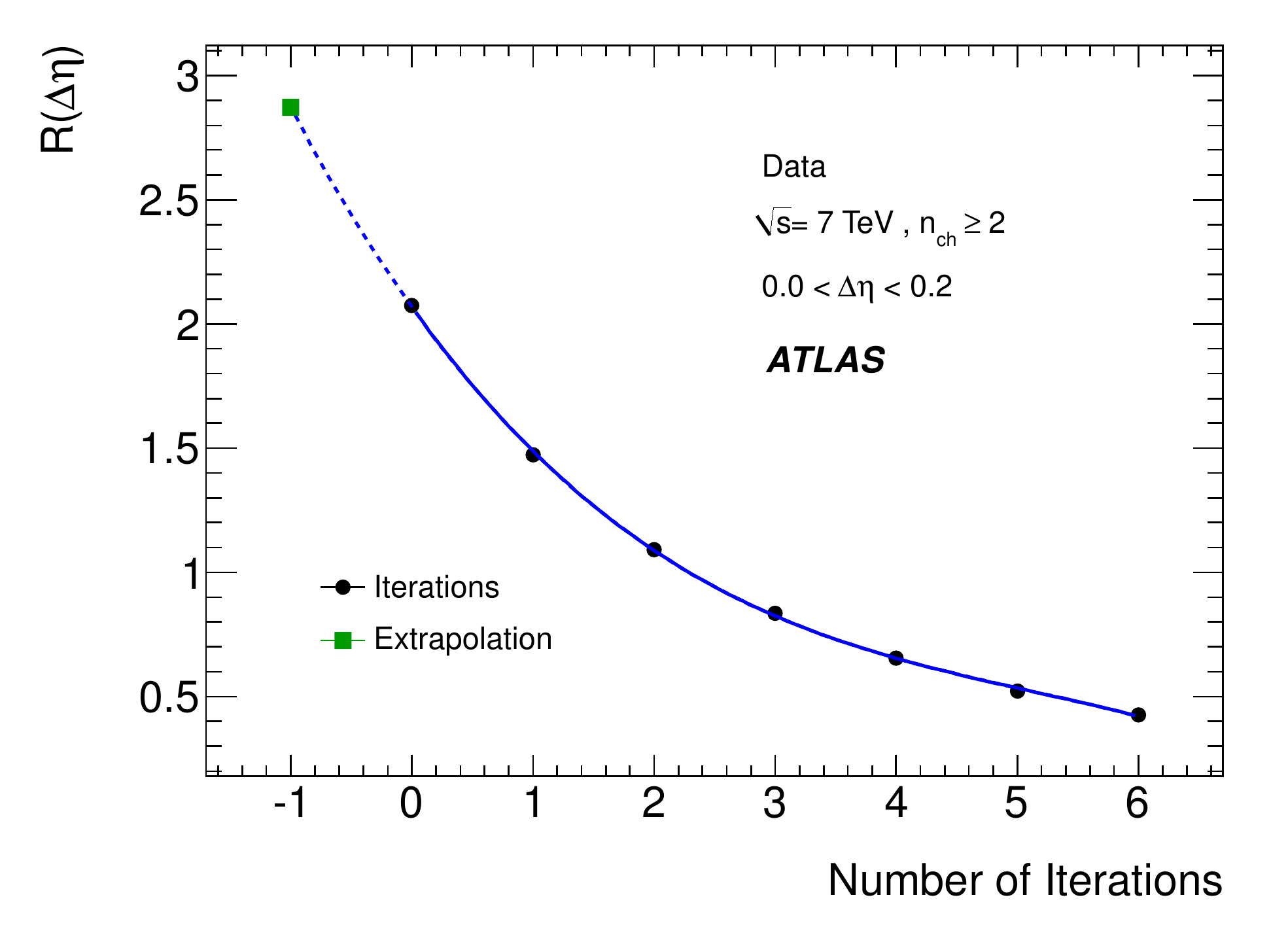}
\put(53,-3){\small(b)}
\end{overpic}
\caption{(a) Two-particle pseudorapidity correlation function, $R(\Delta\eta)$, after each iteration of the probabilistic track removal correction method. (b) Example of a third-degree polynomial fit to the values of $R(\Delta\eta)$ in a central bin.  Plots of $R(\Delta\eta)$ shown here and throughout this paper are symmetric around $\Delta\eta$ = 0 by construction.}
\label{fig:cmethod7TeV}
\end{center}
\end{figure}

The effect of tracks lost due to inefficiencies in the reconstruction is corrected for using the model-independent Hit Backspace Once More (HBOM) method described in ref.~\cite{Monk:2011pg}. Figure \ref{fig:cmethod7TeV} illustrates this correction procedure, in which the observable is first computed using all reconstructed tracks that satisfy the analysis selection criteria (solid circles in figure \ref{fig:cmethod7TeV}a). The track reconstruction efficiency as a function of the track $\pT$ and $\eta$ is taken from ref.~\cite{MB2:2011}.  For each track, a random number, $r_{i}$, uniformly distributed between zero and one, is generated and compared to the track reconstruction efficiency, $\epsilon_{i}(\pT,\eta)$. A track is removed from the sample if the random number is greater than the track reconstruction efficiency. This results in a subset of the original tracks, defined by all those tracks satisfying

\begin{equation}
\epsilon_{i}(\pT,\eta) > r_{i}.
\end{equation}

Tracks removed from the sample in this way are lost with the same probability as they would be lost by the detector reconstruction.  The reduced  sample of tracks, which now expresses an exaggerated detector reconstruction effect, is used to compute the observable again. This defines one iteration of the track-removal procedure (the full set of uncorrected tracks is the 0th iteration). A second iteration takes as input the tracks used in the first iteration and again uses a new set of random numbers to remove additional tracks according to their track reconstruction efficiency. The observable is then determined from this third sample of tracks that shows a larger detector effect than the first two samples. Further iterations are carried out in the same way.

Using this probabilistic track removal method, it is possible to quantify the effect that detector inefficiencies have on an observable; each iteration corresponds to an additional application of the detector effect to the data. The value of each bin of the observable can be plotted as a function of the iteration number (0, 1, 2, ..., $N$) and a function can be fitted to the resulting set of points. By extrapolating the fit to the $N$=$-1$ iteration, an estimate of the observable at the particle level, without detector effects, can be made. In this analysis the fit and subsequent extrapolation use a third-degree polynomial, with six track removal iterations in total used as the input, as illustrated in figure \ref{fig:cmethod7TeV}b for one particular $\Delta\eta$ bin.

\begin{figure}[t]
\begin{center}
\begin{overpic}[height=70mm]{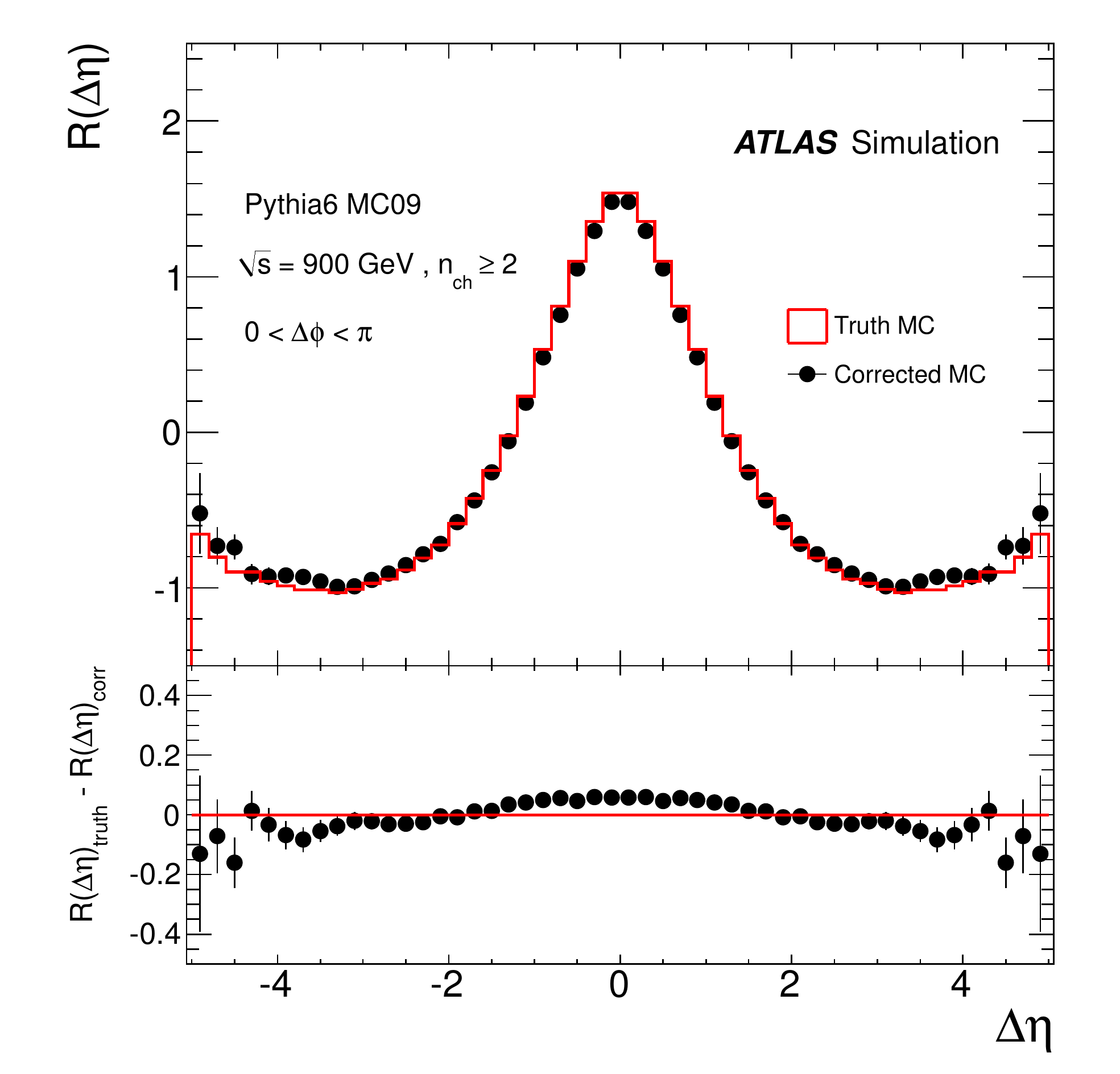}
\put(52,-3){\small(a)}
\end{overpic} \\
\vspace{5mm}
\begin{overpic}[height=70mm]{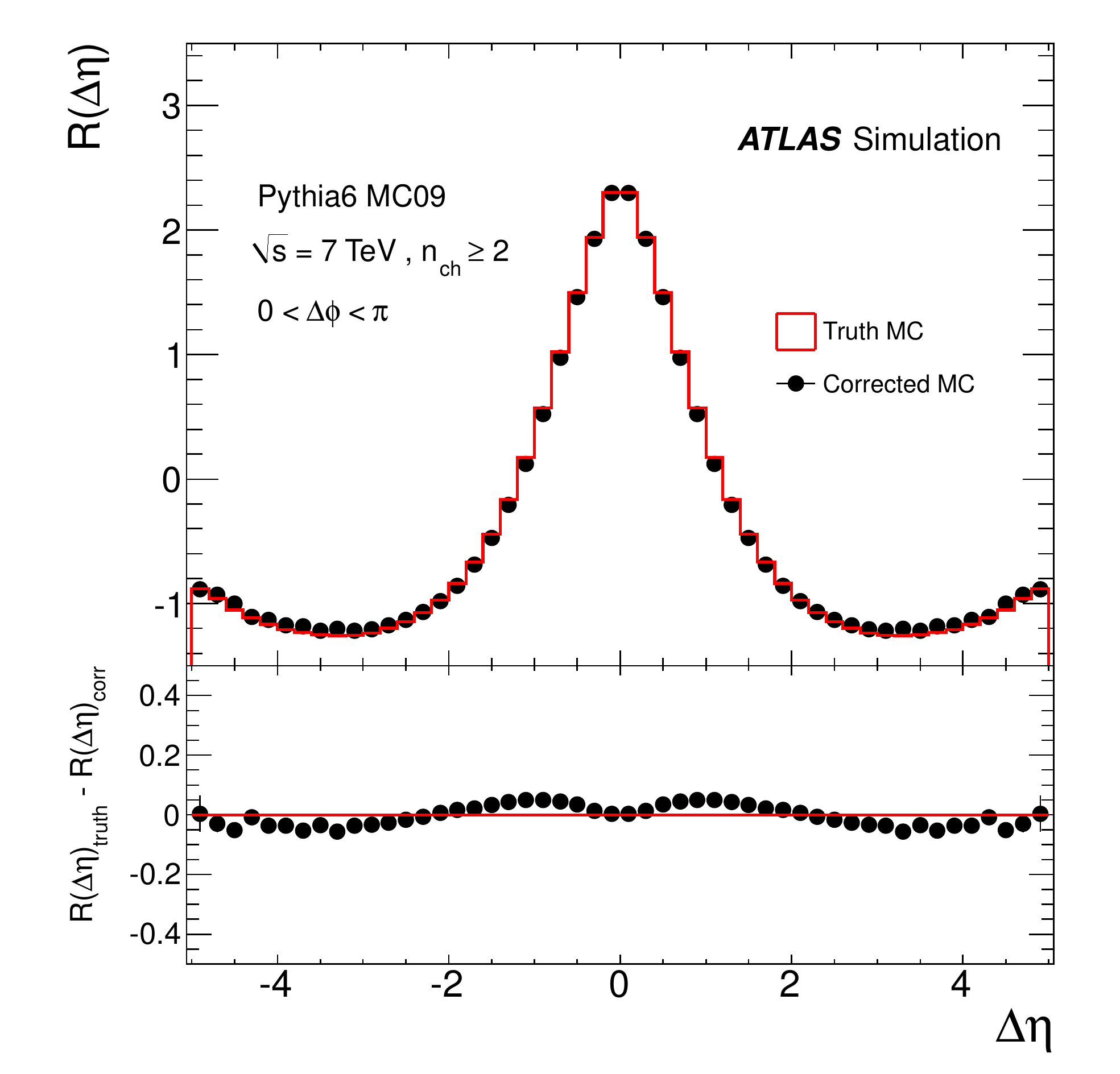}
\put(52,-3){\small(b)}
\end{overpic}
\begin{overpic}[height=70mm]{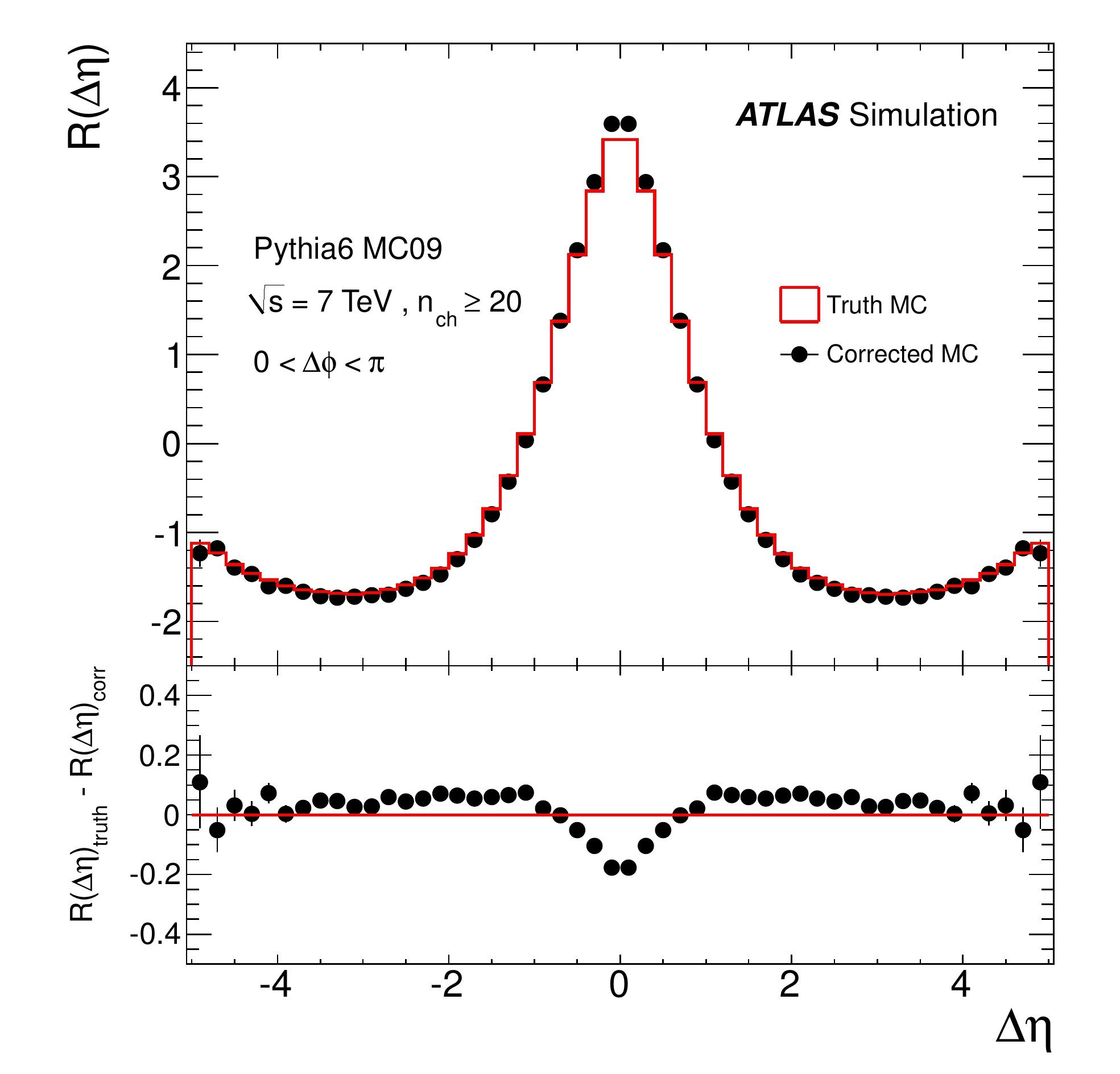}
\put(52,-3){\small(c)}
\end{overpic}
\caption{Comparison between the generated and corrected two-particle correlation functions $R(\Delta\eta)$ at (a) $\sqrt{s} =$ 900 GeV and 7 TeV for the phase-space regions (b) $n_{ch} \geq 2$ and (c) $n_{ch} \geq 20$ for the Monte Carlo MC09 tune. The absolute difference between the generated and corrected MC as a function of $\Delta\eta$ is shown underneath each distribution.}
\label{fig:closure}
\end{center}
\end{figure}

No specific correction is made for tracks arising from the secondary decays of longer lived particles nor from material interactions. The effect of secondary particles is discussed in section \ref{sec:systClosure}.

In order to test the correction procedure, Monte Carlo studies are performed in which the distributions at the generated particle level are recovered from a fully detector-simulated and reconstructed Monte Carlo sample. This is shown in figure \ref{fig:closure} for the \pythia\ 6 MC09 tune at $\sqrt{s} =$ 900 GeV and 7 TeV for $R(\Delta\eta)$ with $0 < \Delta\phi < \pi$. The size of the correction that must be applied to the uncorrected correlation function depends on the value of $\Delta\eta$ and $\Delta\phi$; in general the correction has an absolute value of around 0.5.  At high $\left|\Delta\eta\right|$ the correction is negative, while at low $\left|\Delta\eta\right|$ the correction is positive, so the overall effect is that the peak at $\Delta\eta=0$ is narrowed significantly.  The generated and corrected distributions are generally in good agreement with each other, showing an absolute difference of approximately 0.05 in all but a few bins. The remaining differences are considered as a systematic uncertainty, as discussed in section \ref{sec:systClosure}. 

All the distributions presented in this paper, both one- and two-dimensional, have been corrected using this procedure. No correction is made for the effects of angular smearing because the resolution of the tracking detector is much better than the bin size used. Any such correction would be insignificant in comparison to the uncertainty on the efficiency correction that is applied.

\section{Statistical and systematic uncertainties}
\label{sec:systematics}

The sources of uncertainty identified for this measurement are described in this section. The most significant contributions to both the one- and two-dimensional distributions are associated with the track reconstruction efficiency and with the correction procedure (arising from the small differences between the generated and corrected Monte Carlo distributions).

\subsection{Statistical uncertainty}

Each of the parameters in the third-degree polynomial fits in the probabilistic track removal method carries a statistical error. The statistical uncertainty in the corrected values is the result of propagating these uncertainties, using the full covariance matrix, when evaluating the third-degree equation at $N$=$-1$. In general, this uncertainty is negligible compared to the ones associated to the track reconstruction efficiency and the non-closure of the correction method.  The exceptions to this are the highest $\Delta\eta$ bins where, due to reduced statistics, the uncertainty on the fit parameters is of the same magnitude as the other uncertainties.

\subsection{Uncertainties on the efficiencies}

In order to evaluate the systematic uncertainty that is associated with uncertainties in the trigger, vertex and track reconstruction efficiencies, the complete analysis is repeated varying each of these quantities, one at a time, according to their uncertainties.  It is assumed there is no difference in the size of these uncertainties between the different collision energies, as discussed in ref.~\cite{MB2:2011}.

The systematic uncertainties associated with the track reconstruction efficiency come from various sources: material effects, track selection, momentum resolution and badly measured high-$\pT$ tracks. The tracking efficiency is obtained in ref.~\cite{MB2:2011} by matching MC generated  charged particles to tracks reconstructed after the full detector simulation.  Uncertainties on the efficiency are determined by varying the parameters of the detector simulation.   These uncertainties are summarised in table \ref{tab:systTrack}. All of these uncertainties are added in quadrature to obtain the total uncertainty on the tracking efficiency in each ($\pT$,$\eta$) bin.

In the case of the trigger and vertex reconstruction efficiencies, the systematic uncertainties are of the order of 1\% for events with a track multiplicity of two and decrease rapidly as the multiplicity increases.

\begin{table}[t]
\begin{tabular}{|l|l|}
\hline \hline
Source					& Uncertainty	\\
\hline
Material description in MC	& Increases at high-$|\eta|$ and decreases at high-$\pT$; \\
						& max. 15\% in the range $2.3 < |\eta| < 2.5$ \\
						& and 100 MeV $< \pT < 150$ MeV	\\
\hline
Track selection				& 1\% in all ($\pT$,$\eta$) bins	 \\
\hline
$\pT$ resolution in MC		& 5\% in the first $\pT$ bin (100 MeV $< \pT < 150$ MeV)	\\
\hline
High-$\pT$ tracks			& 10\% due to the requirement on the track-fit $\chi^{2}$ probability \\
\cline{2-2}
						& $\eta$ and $\pT$ dependent uncertainty due to \\ 
						& mis-measured tracks	\\
\hline \hline
\end{tabular}
\centering
\caption{Uncertainties associated to the track reconstruction efficiency. Values are taken from and are identical to those in ref.~\cite{MB2:2011}, which provides a more complete discussion on this subject.}
\label{tab:systTrack}
\end{table}

\subsection{Uncertainty due to lack of closure using Monte Carlo samples}
\label{sec:systClosure}

The absolute difference between the generated and corrected Monte Carlo distributions is used as the systematic uncertainty resulting from the small non-closure of the correction procedure. From this it is possible to derive an estimate for the non-closure uncertainty in data as a function of the corrected value of $R$.

\begin{figure}[t]
\begin{center}
\includegraphics[height=80mm]{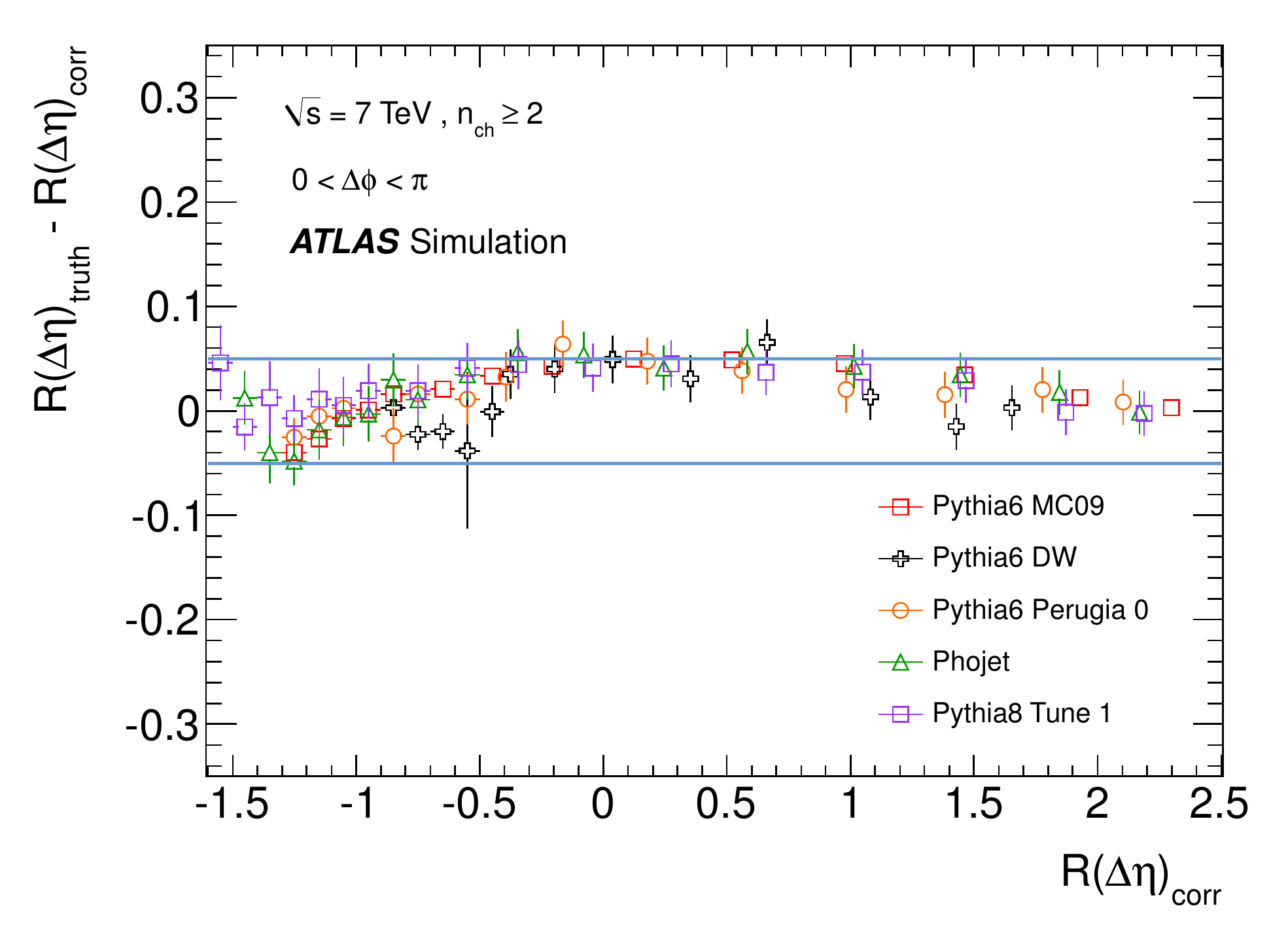}
\caption{Difference between the generated and corrected two-particle correlation functions as a function of $R(\Delta\eta)_{corr}$ for various Monte Carlo tunes at $\sqrt{s} =$ 7 TeV for $n_{ch}\geq2$.}
\label{fig:closureSyst7}
\end{center}
\end{figure}

This is illustrated in figure \ref{fig:closureSyst7} for $R(\Delta\eta)$ (where $\Delta\phi$ was integrated from 0 to $\pi$) for different tunes at $\sqrt{s} =$ 7 TeV for $n_{ch}\geq2$. The absolute amount of non-closure does not depend strongly on either the Monte Carlo model or on the value of $R(\Delta\eta)$ and, for this example, lies generally in the region between $\pm$0.05.  Therefore in this case a systematic uncertainty of 0.05 is assigned to all bins in the data distribution.

The systematic uncertainty associated with the non-closure of the correction method is calculated in this manner for each of the one-dimensional projections and the two-dimensional distributions. In most cases the systematic uncertainty has a value of 0.05 or smaller and has very little dependence on $R$; however, in a small number of cases for some of the $n_{ch}\geq20$ distributions, the uncertainty can be as large as 0.2.

The small non-closure of the corrected samples with the hadron-level Monte Carlo distributions is primarily caused by the introduction of additional particles in the detector simulation that are not produced by the Monte Carlo generator. Such secondary particles can be produced either by the decays of longer-lived particles that are set stable by the generator (primarily neutral kaons), or by interactions between primary particles and the detector material.\  Both sources of secondary particles produce  correlated pairs of tracks, and as such contribute to the non-closure in the same manner;  however, the pairs of secondary particles initiated by detector interactions produce an effect that is between five and ten times as large as the effect produced by neutral decay products.

Using the reconstruction efficiency to remove tracks from the data sample provides an approximation to the effect that the detector has on the observable.  This approximation does not account for the effect of fake tracks or of tracks that arise from secondary decays and material interactions.  In order to apply a correction for these small effects, the probabilities that such particles are produced would need to be assessed so that they could be used to add  additional particles probabilistically in each iteration of the correction procedure.  Studies using the full detector simulation show that the size of such corrections would be small in comparison to the correction for tracking inefficiencies and would be of a similar size to other systematic uncertainties present in this measurement.  This justifies the approach of  applying a correction for tracking inefficiencies, which is the largest detector effect, and assigning a non-closure systematic based on studies using the full detector simulation.

\section{Results and discussion}
\label{sec:results}

\begin{figure}[t]
\begin{center}
\begin{overpic}[height=70mm]{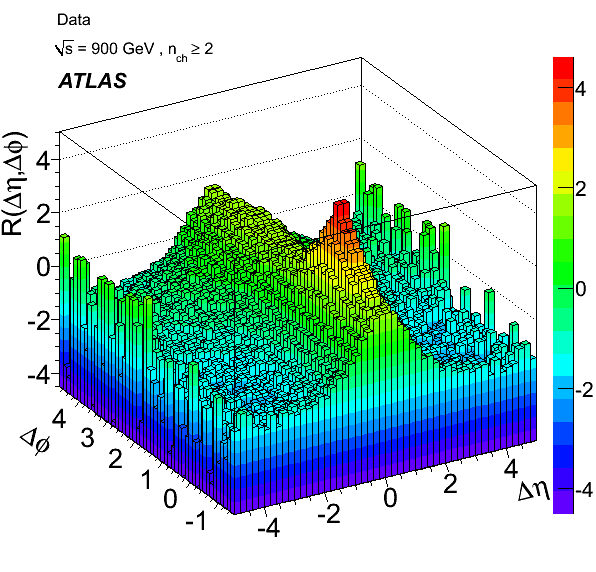}
\put(45,-3){(a)}
\end{overpic}
\begin{overpic}[height=70mm]{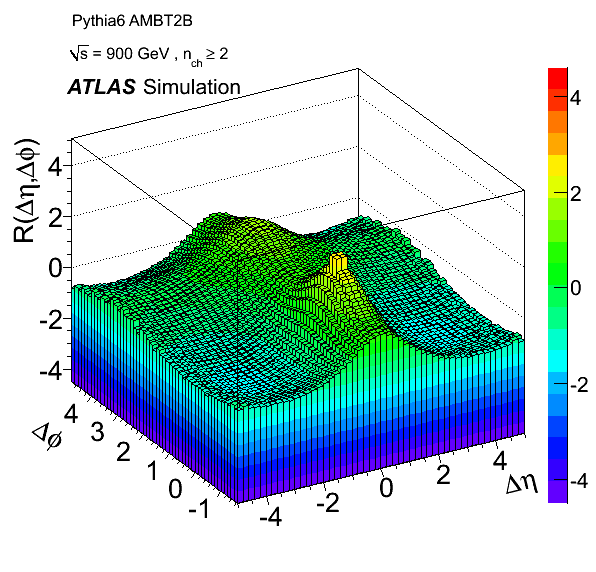}
\put(45,-3){(b)}
\end{overpic}
\caption{Corrected $R(\Delta\eta,\Delta\phi)$ two-particle correlation functions for (a) data and (b) the AMBT2B Monte Carlo tune, at $\sqrt{s} =$ 900 GeV.}
\label{fig:2D900}
\end{center}
\end{figure}

\begin{figure}[t]
\begin{center}
\begin{overpic}[height=70mm]{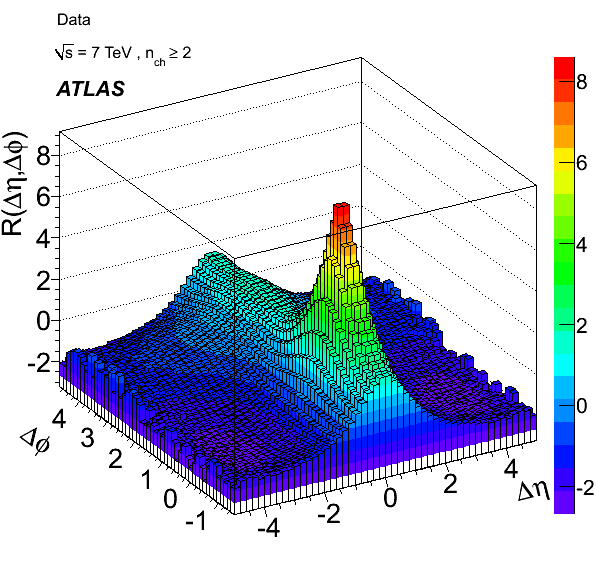}
\end{overpic}
\begin{overpic}[height=72mm]{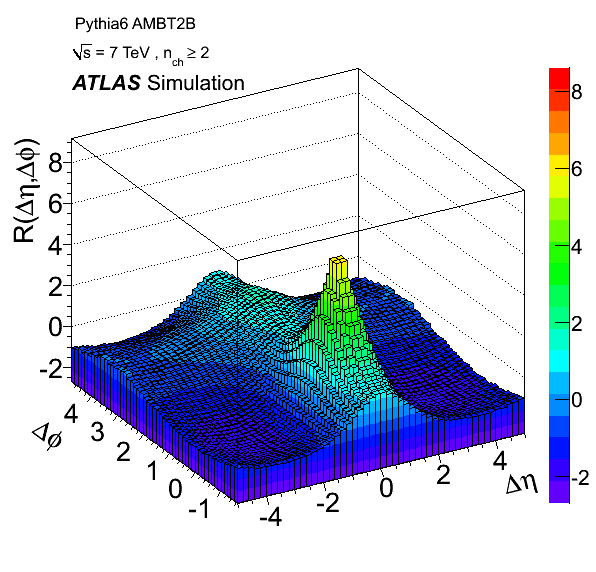}
\end{overpic}
\begin{overpic}[height=70mm]{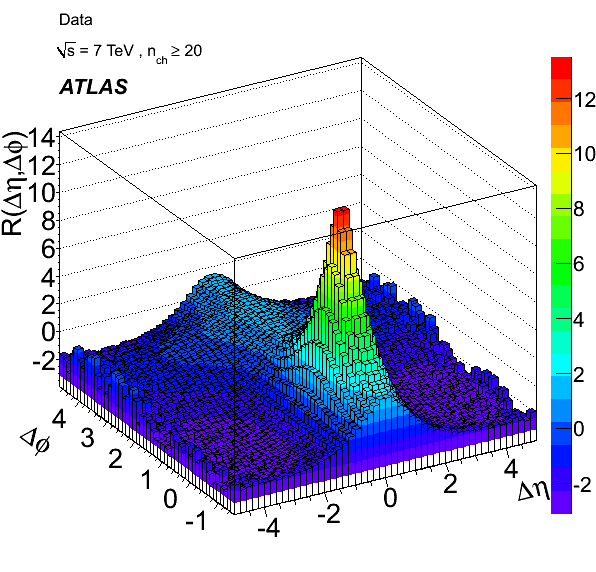}
\put(45,95){(a)}
\put(45,-3){(c)}
\end{overpic}
\begin{overpic}[height=72mm]{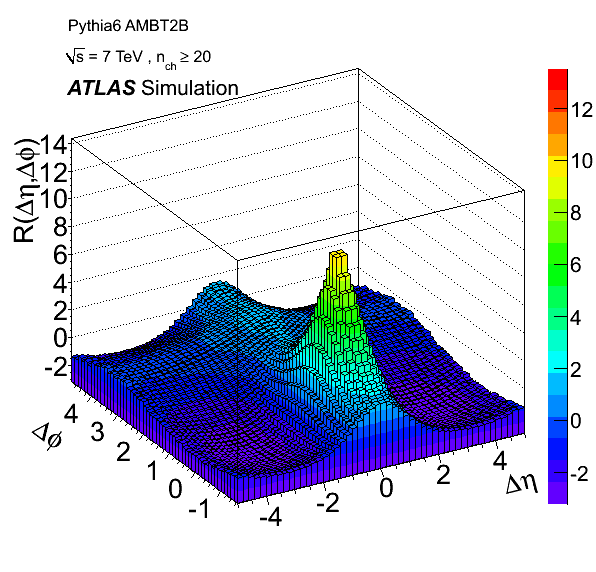}
\put(45,95){(b)}
\put(45,-3){(d)}
\end{overpic}
\caption{Corrected $R(\Delta\eta,\Delta\phi)$ two-particle correlation functions at $\sqrt{s} =$ 7 TeV for $n_{ch}\geq2$ (a) data and (b) Monte Carlo (AMBT2B), and for $n_{ch}\geq20$ (c) data and (d) Monte Carlo (AMBT2B). These plots are symmetric around $\Delta\eta$ = 0 by construction.}
\label{fig:2D7}
\end{center}
\end{figure}

The corrected inclusive two-particle angular correlation functions for data and the \pythia\ 6 AMBT2B Monte Carlo tune are shown in figures \ref{fig:2D900} and \ref{fig:2D7} for $\sqrt{s} =$ 900 GeV and 7~TeV, respectively. 

At both energies a complex structure can be observed across the full $\Delta\eta$ and $\Delta\phi$ range. Several components can be identified that reflect the contributions from different underlying processes to the correlation structure. Most obvious are the near-side correlations,  which are seen as a sharp peak around (0,0) that can be attributed  to particles originating from the same (relatively) high-$\pT$ process.  Such closely correlated particles indicate the  emergence of jet-like structures within this minimum bias event selection. Another prominent structure in the two-dimensional distribution is the away-side correlation, which is seen as a ridge extending across the whole of the $\Delta\eta$ range near $\Delta\phi \sim \pi$.  This away-side activity arises from the recoil of one particle against another and would be seen quite strongly in, for example, back-to-back jet production.  A third distinct structure  is the broad Gaussian-like structure spanning the $\Delta\phi$ domain at $\Delta\eta=0$ with a width of $\Delta\eta\simeq2$.  This last structure can be related to processes such as the decays of resonances, clusters or string fragmentation.

As the centre-of-mass energy increases, so too does the height of the peak at (0,0).  This peak becomes even more pronounced for the set of events with a higher charged-particle multiplicity (figure \ref{fig:2D7}c). Conversely, the height of the away-side ridge is constant regardless of the collision energy or multiplicity. \pythia~6 AMBT2B exhibits similar structures to those observed in data; however it does not reproduce the strength of the correlations.

\subsection{Projections in $\Delta\eta$ and $\Delta\phi$}

In all of the following figures, the solid markers correspond to the corrected data values, the error bars are statistical only (after correcting for track detection inefficiency) and the solid bands correspond to the total uncertainty in the bin, which is obtained by adding in quadrature the statistical and systematic uncertainties described in section \ref{sec:systematics}. The data distributions are compared to different Monte Carlo tunes. In general, none of the models reproduce the strength of the correlations seen in data. Some approximate the shape of the distributions more closely than others, but for all distributions \herwig++ is the most discrepant with the data.

\begin{figure}[t]
\begin{center}
\begin{overpic}[height=70mm]{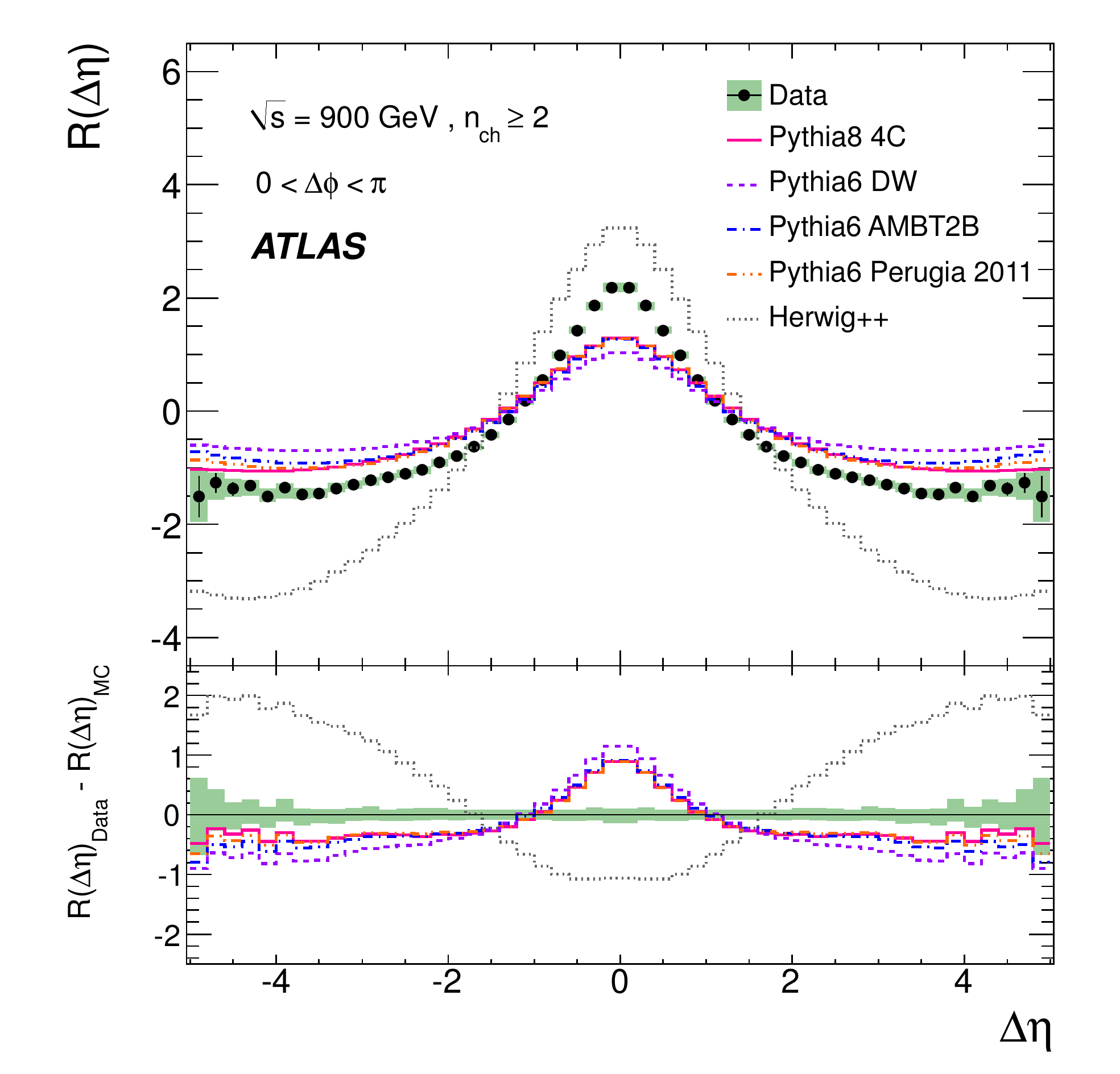}
\put(85,85){(a)}
\end{overpic} \\
\begin{overpic}[height=70mm]{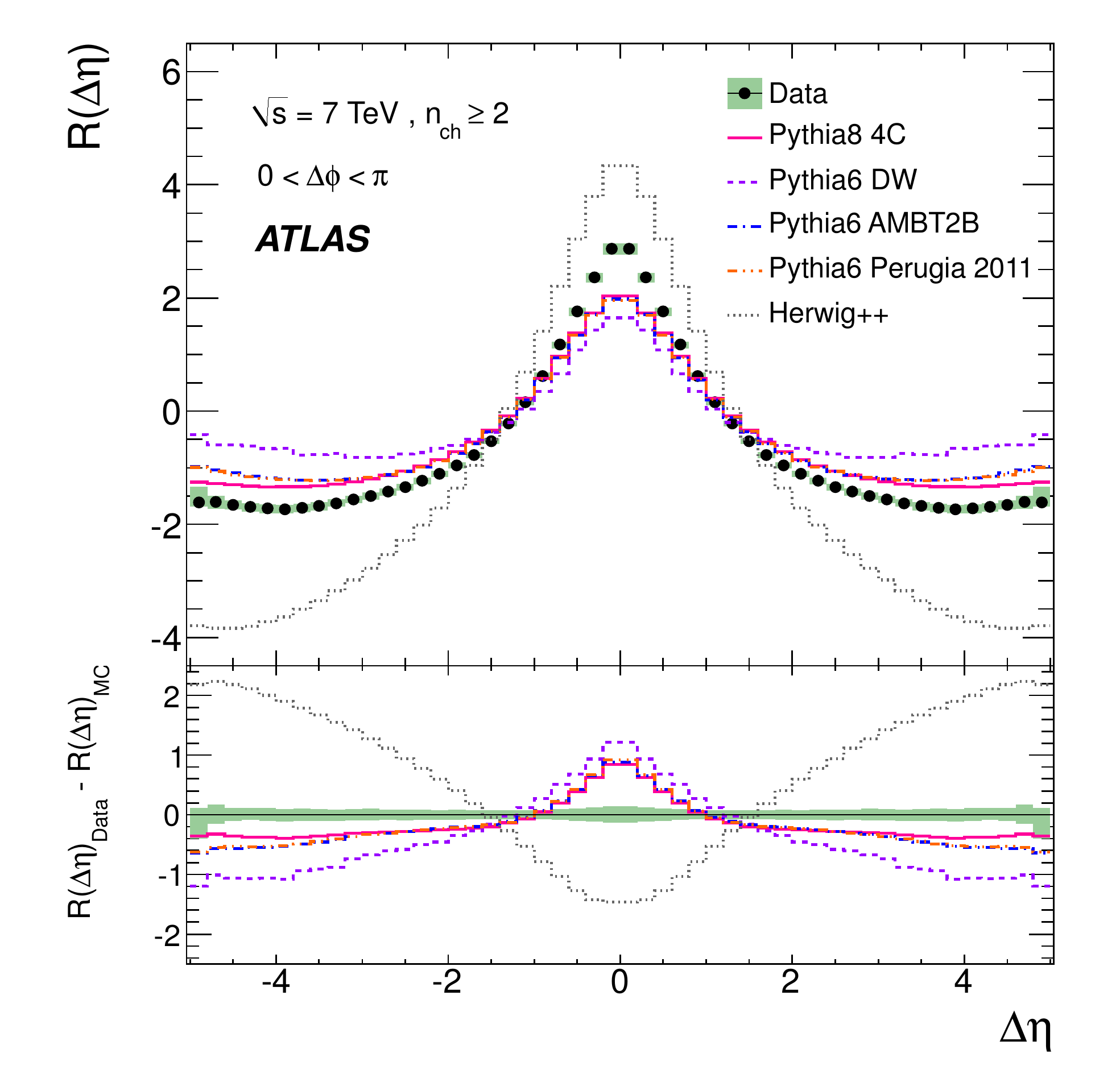}
\put(85,85){(b)}
\end{overpic}
\begin{overpic}[height=70mm]{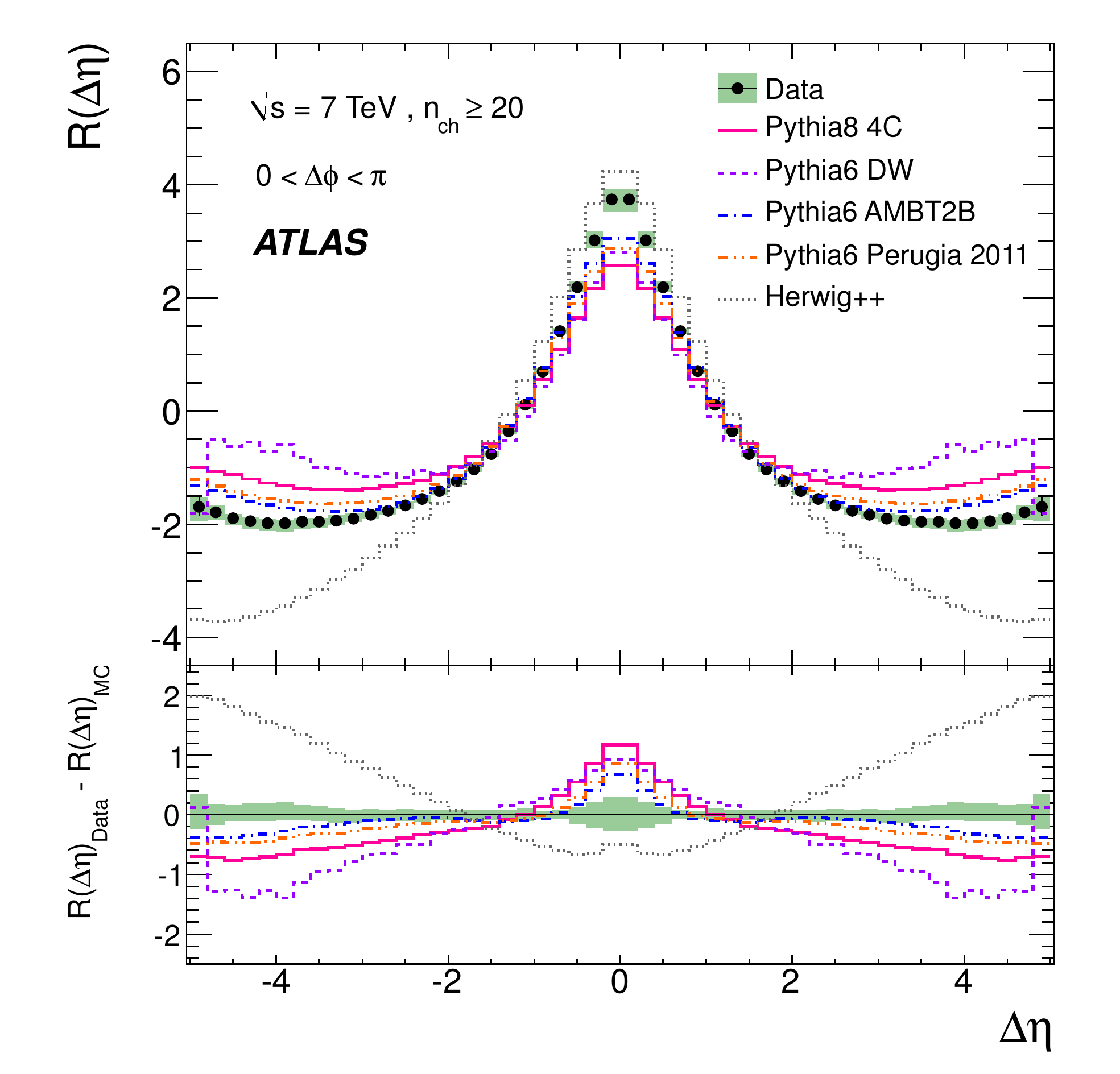}
\put(85,85){(c)}
\end{overpic}
\caption{Corrected $R(\Delta\eta$) two-particle correlation functions obtained by integrating the foreground and background distributions over $\Delta\phi$ between 0 and $\pi$ for data and several different Monte Carlo tunes at (a) $\sqrt{s} =$ 900 GeV and 7 TeV for events with a charged-particle multiplicity (b) $n_{ch}\geq2$ and (c) $n_{ch}\geq20$.}
\label{fig:deta0pi}
\end{center}
\end{figure}

The corrected pseudorapidity correlation distributions are shown in figure \ref{fig:deta0pi} and were obtained by integrating separately the foreground and background functions over $\Delta\phi$ between 0 and $\pi$. The AMBT2B and Perugia~2011 tunes of \pythia~6 together with tune 4C of \pythia~8 show the best agreement with the data distributions for events with $n_{ch}\geq2$ at both collision energies.  For the sample of events satisfying $n_{ch}\geq20$ the AMBT2B tune shows the closest agreement to the data over the entire $\Delta\eta$ range, although the agreement cannot be considered satisfactory.

\begin{figure}[t]
\begin{center}
\begin{overpic}[height=70mm]{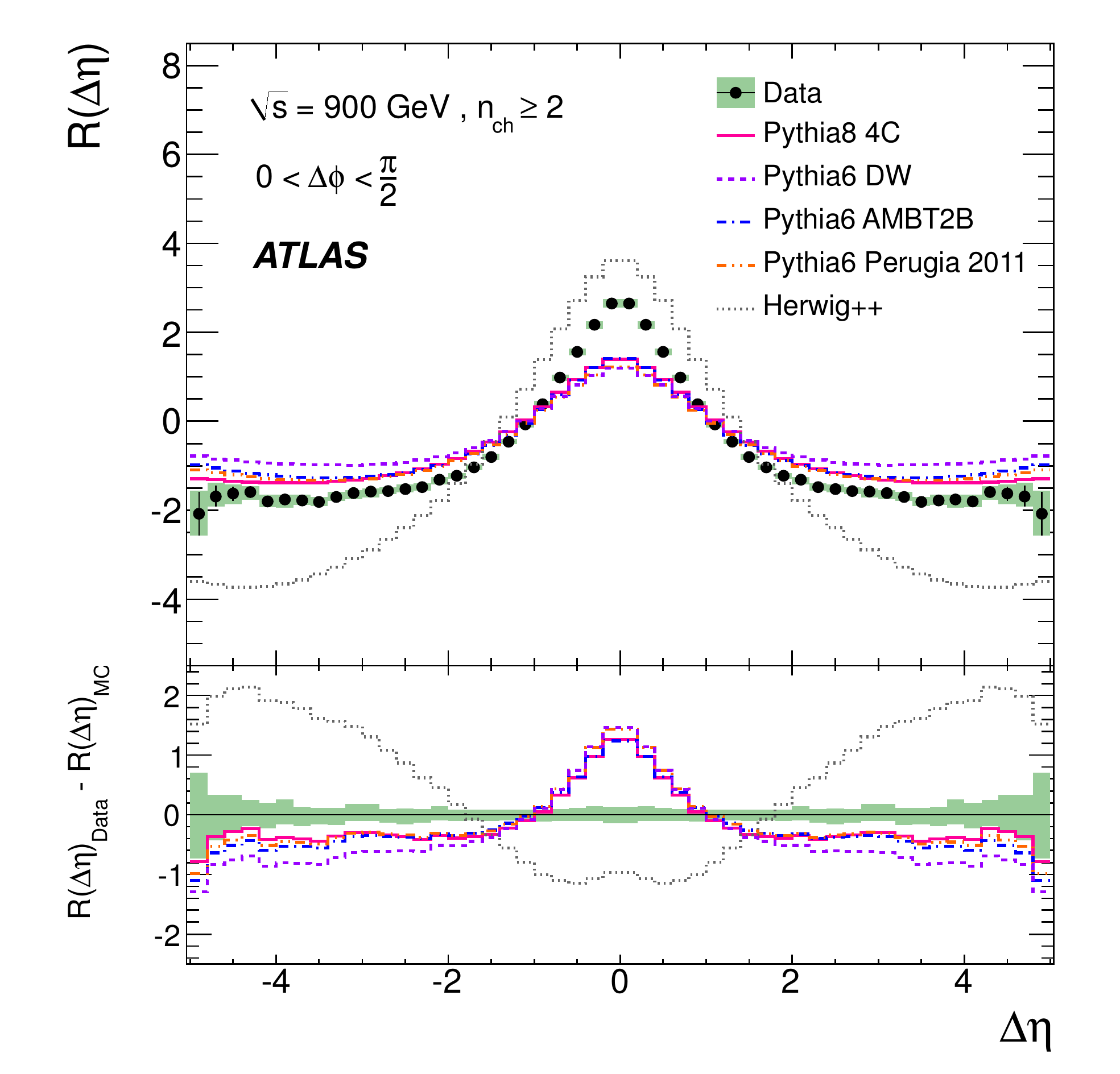}
\put(85,85){(a)}
\end{overpic} \\
\begin{overpic}[height=70mm]{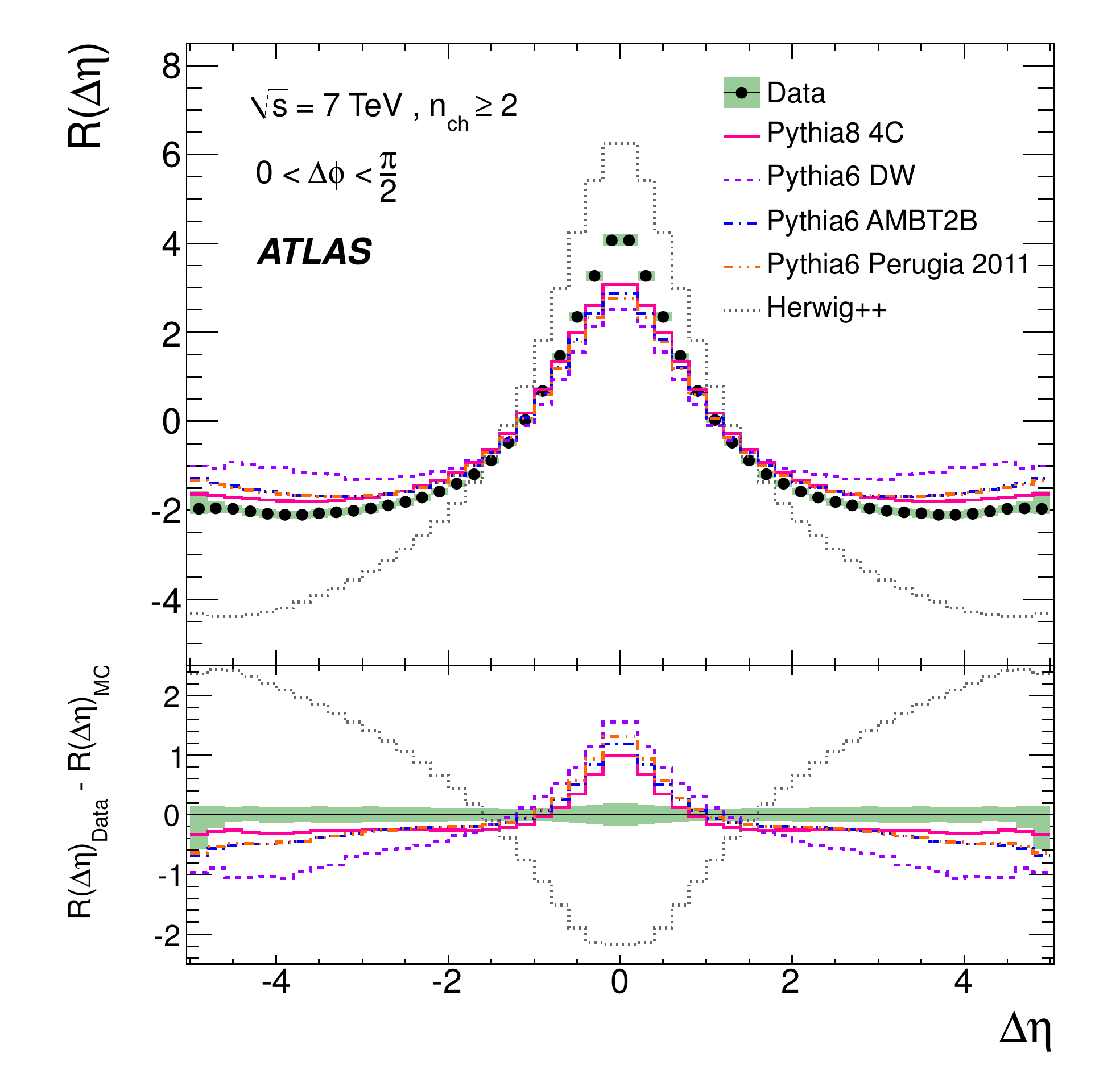}
\put(85,85){(b)}
\end{overpic}
\begin{overpic}[height=70mm]{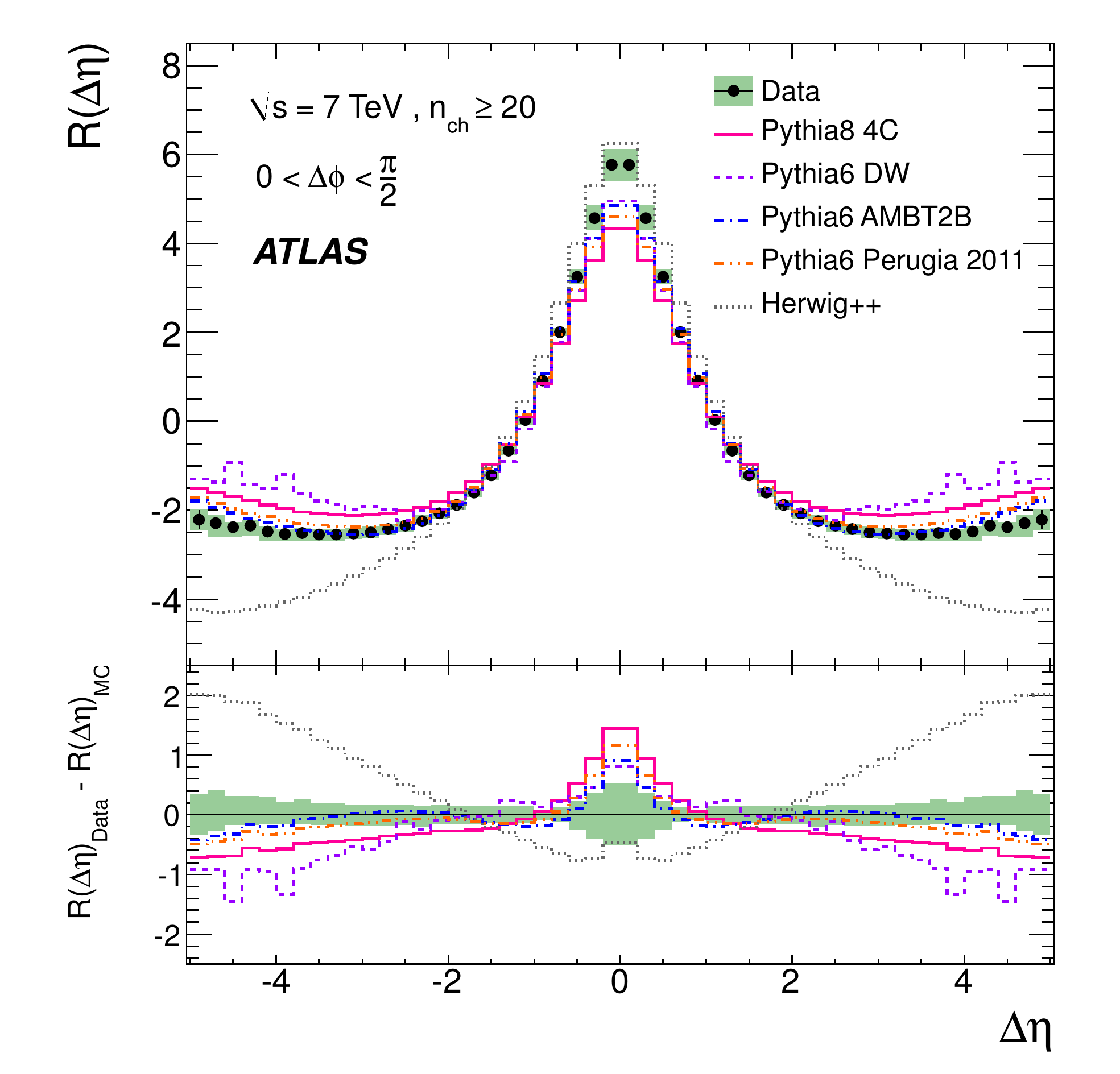}
\put(85,85){(c)}
\end{overpic}
\caption{Corrected $R(\Delta\eta$) two-particle correlation functions obtained by integrating the foreground and background distributions over $\Delta\phi$ between 0 and $\frac{\pi}{2}$ for data and the different Monte Carlo tunes at (a) $\sqrt{s} =$ 900 GeV and 7 TeV for events with a charged-particle multiplicity (b) $n_{ch}\geq2$ and (c) $n_{ch}\geq20$.}
\label{fig:deta0pi2}
\end{center}
\end{figure}

In order to separate the $\eta$ correlations on the near side ($\Delta\phi\simeq0$) from those on the away side ($\Delta\phi\simeq\pi$), the pseudorapidity correlation function was also calculated by integrating the foreground and background functions over $\Delta\phi$ within the more limited ranges of  $\big(0:\frac{\pi}{2}\big)$ (near side) and $\big(\frac{\pi}{2}:\pi\big)$ (away side).  Figure \ref{fig:deta0pi2} shows that, when only the near side is included, the peak at (0, 0) becomes narrower, higher and more pronounced, which indicates a stronger correlation between nearby particles.\ Tune 4C of \pythia~8 provides the best description of the data across the entire $\Delta\eta$ range for the $n_{ch}\geq2$ event samples at both energies, while the AMBT2B tune provides the most reasonable description of events for which $n_{ch}\geq20$.

\begin{figure}[t]
\begin{center}
\begin{overpic}[height=70mm]{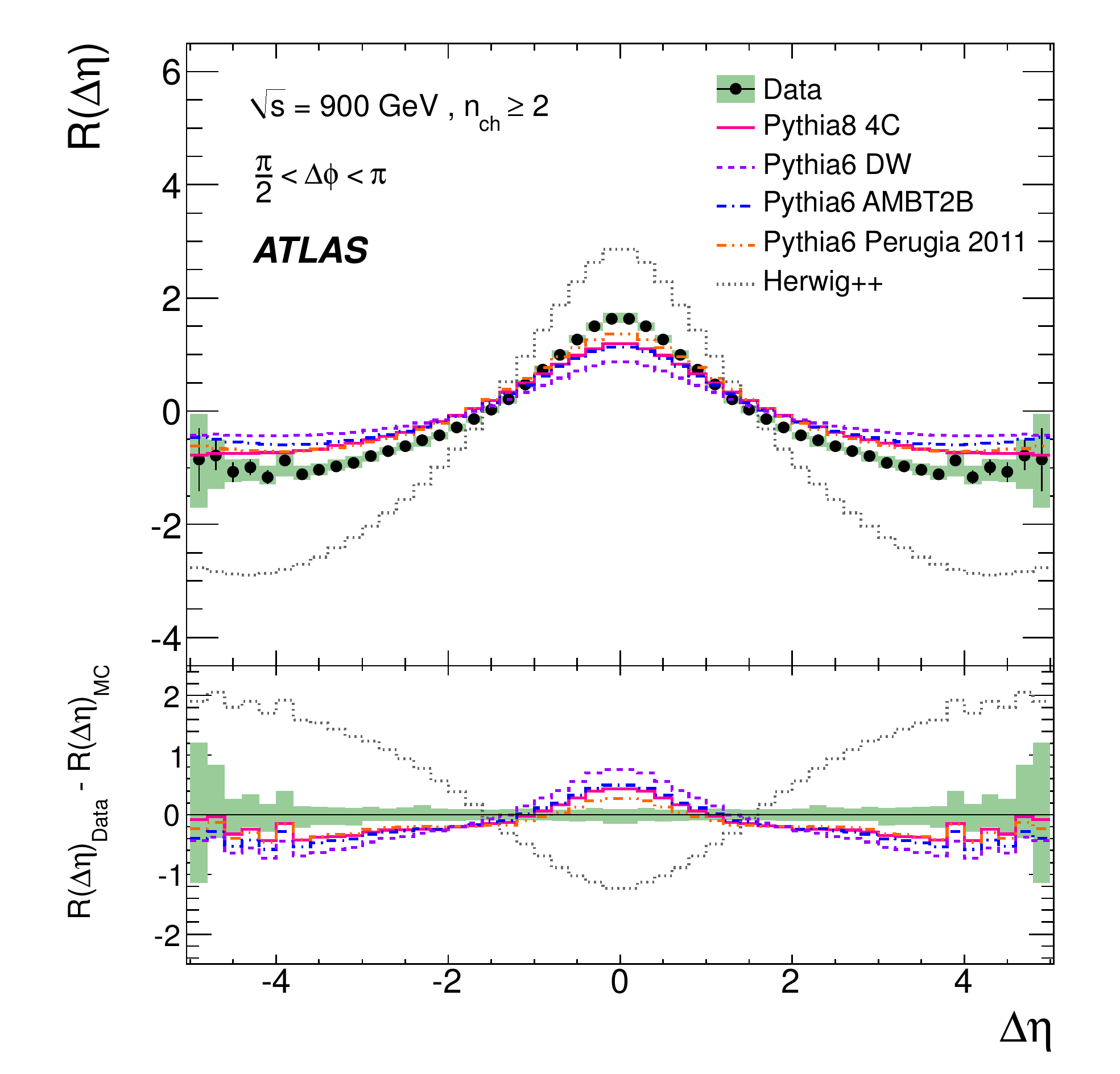}
\put(85,85){(a)}
\end{overpic} \\
\begin{overpic}[height=70mm]{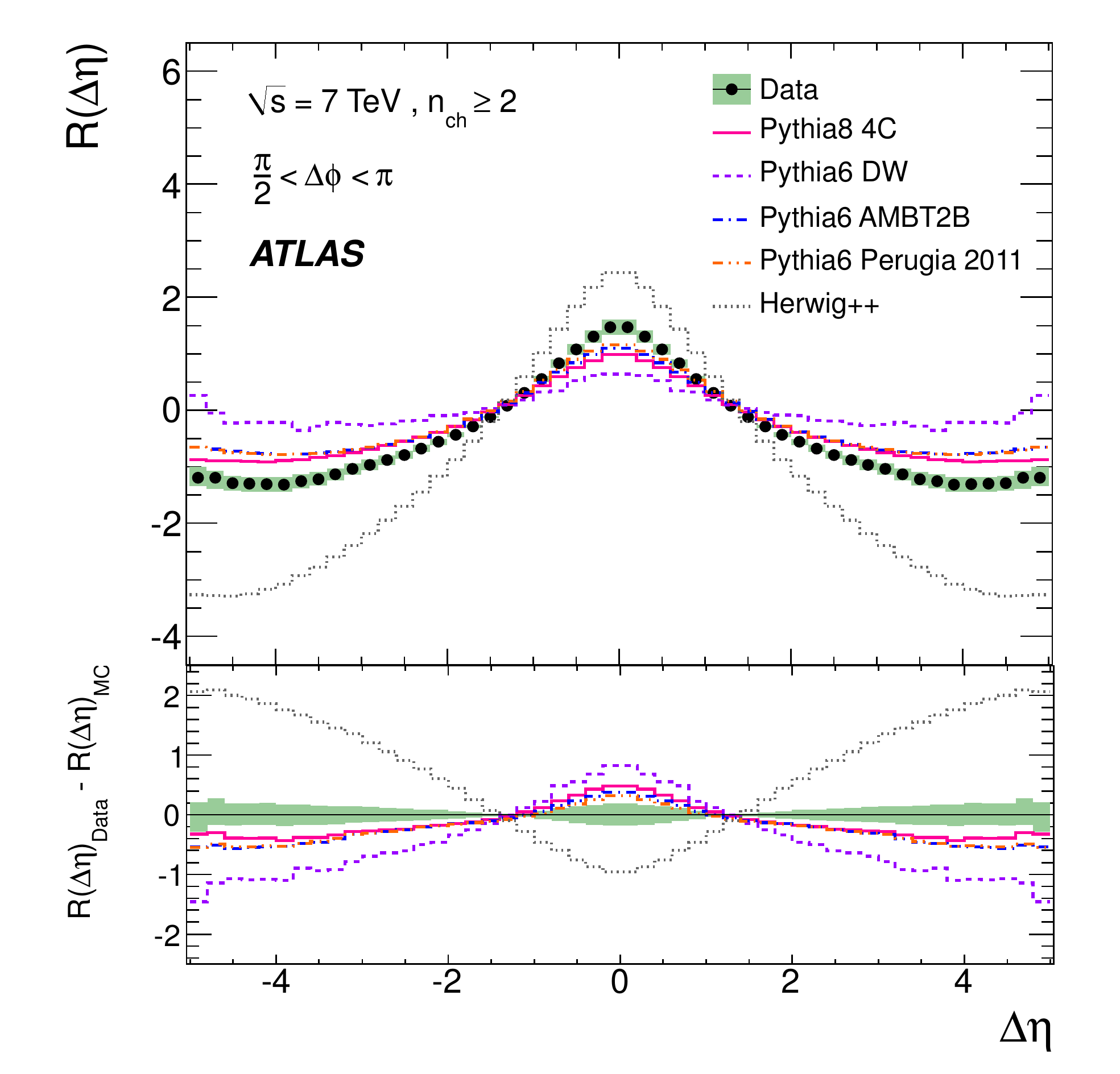}
\put(85,85){(b)}
\end{overpic}
\begin{overpic}[height=70mm]{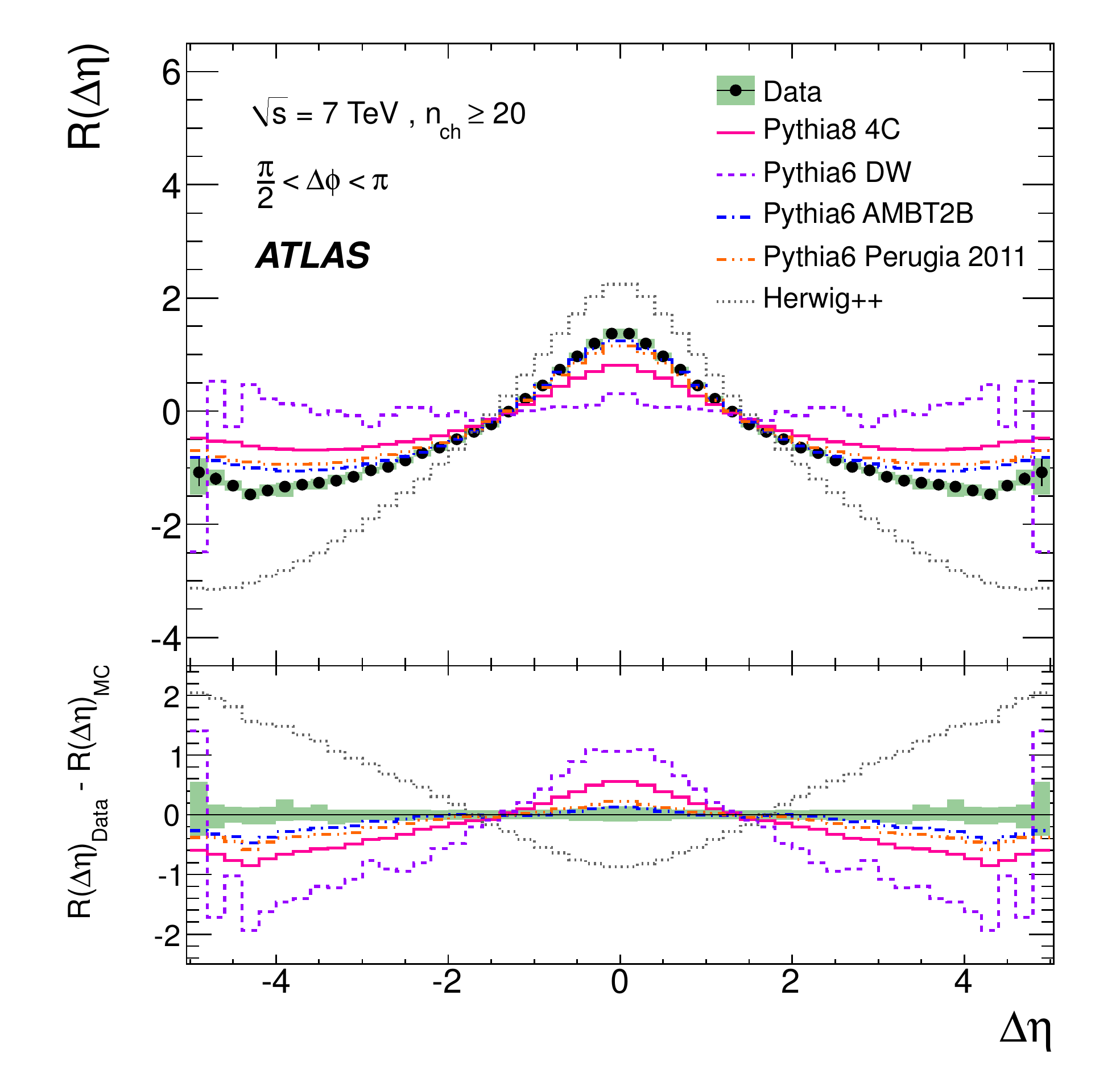}
\put(85,85){(c)}
\end{overpic}
\caption{Corrected $R(\Delta\eta$) two-particle correlation functions obtained by integrating the foreground and background distributions over $\Delta\phi$ between $\frac{\pi}{2}$ and $\pi$ for data and the different Monte Carlo tunes at (a) $\sqrt{s} =$ 900 GeV and 7 TeV for events with a charged-particle multiplicity (b) $n_{ch}\geq2$ and (c) $n_{ch}\geq20$.}
\label{fig:detapi2pi}
\end{center}
\end{figure}

The $\eta$ correlations on the away side are shown in figure  \ref{fig:detapi2pi}, and here most of the tunes (with the exception of \herwig++) display a better agreement with data.  The Perugia~2011 tune shows a particularly good agreement with data across the $\Delta\eta$ range at both collision energies.

\begin{figure}[t]
\begin{center}
\begin{overpic}[height=70mm]{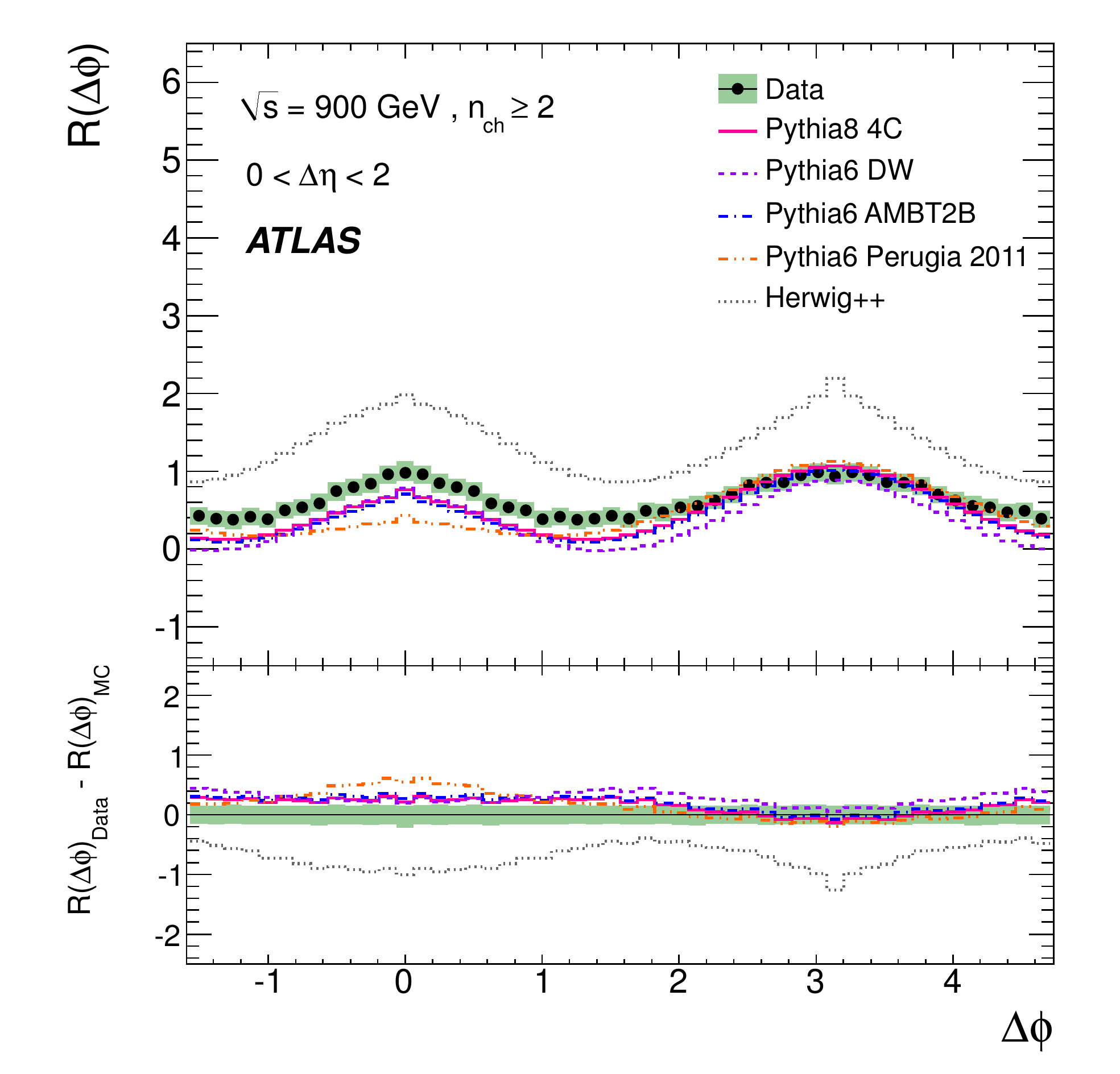}
\put(85,85){(a)}
\end{overpic} \\
\begin{overpic}[height=70mm]{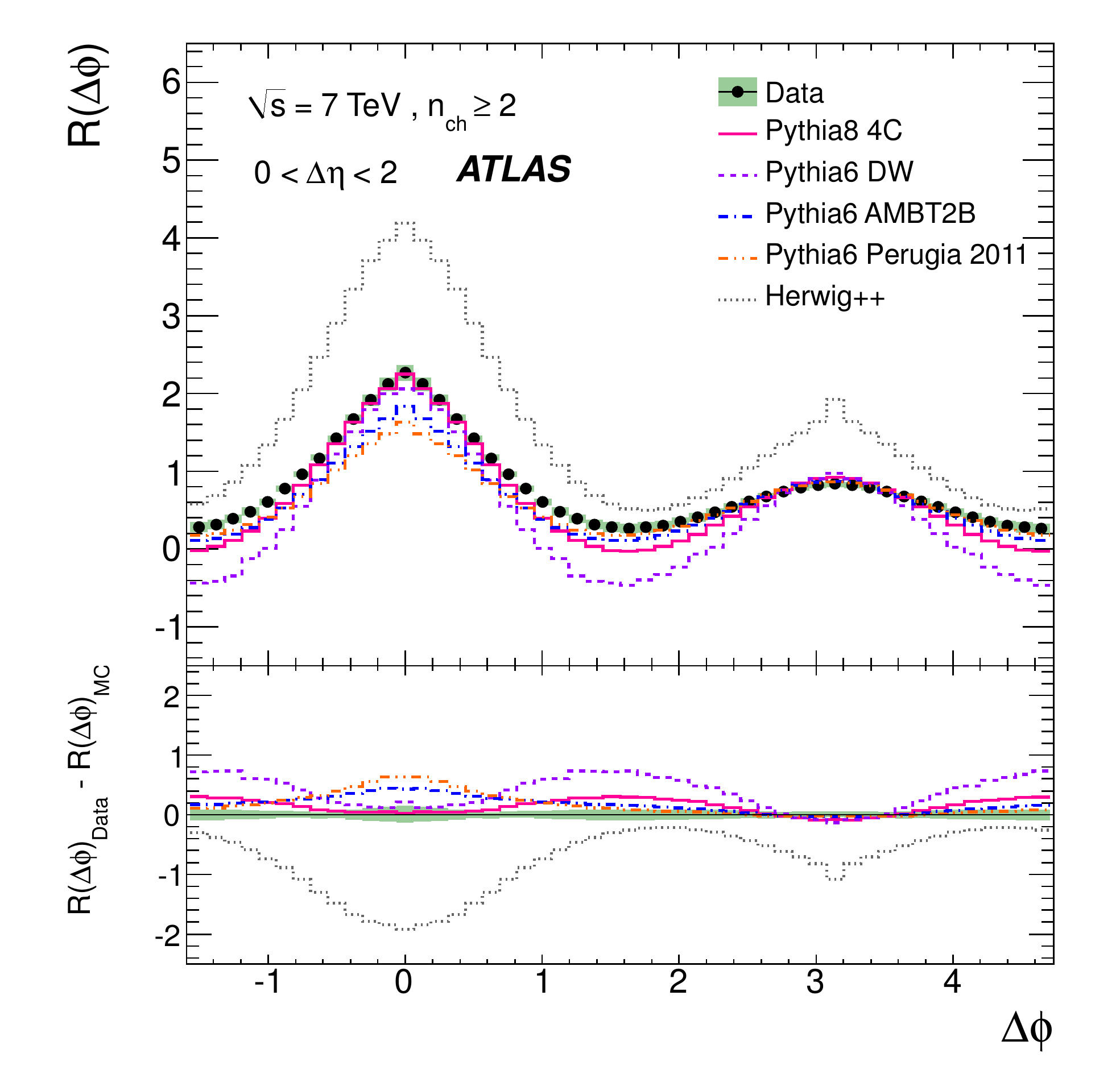}
\put(85,85){(b)}
\end{overpic}
\begin{overpic}[height=70mm]{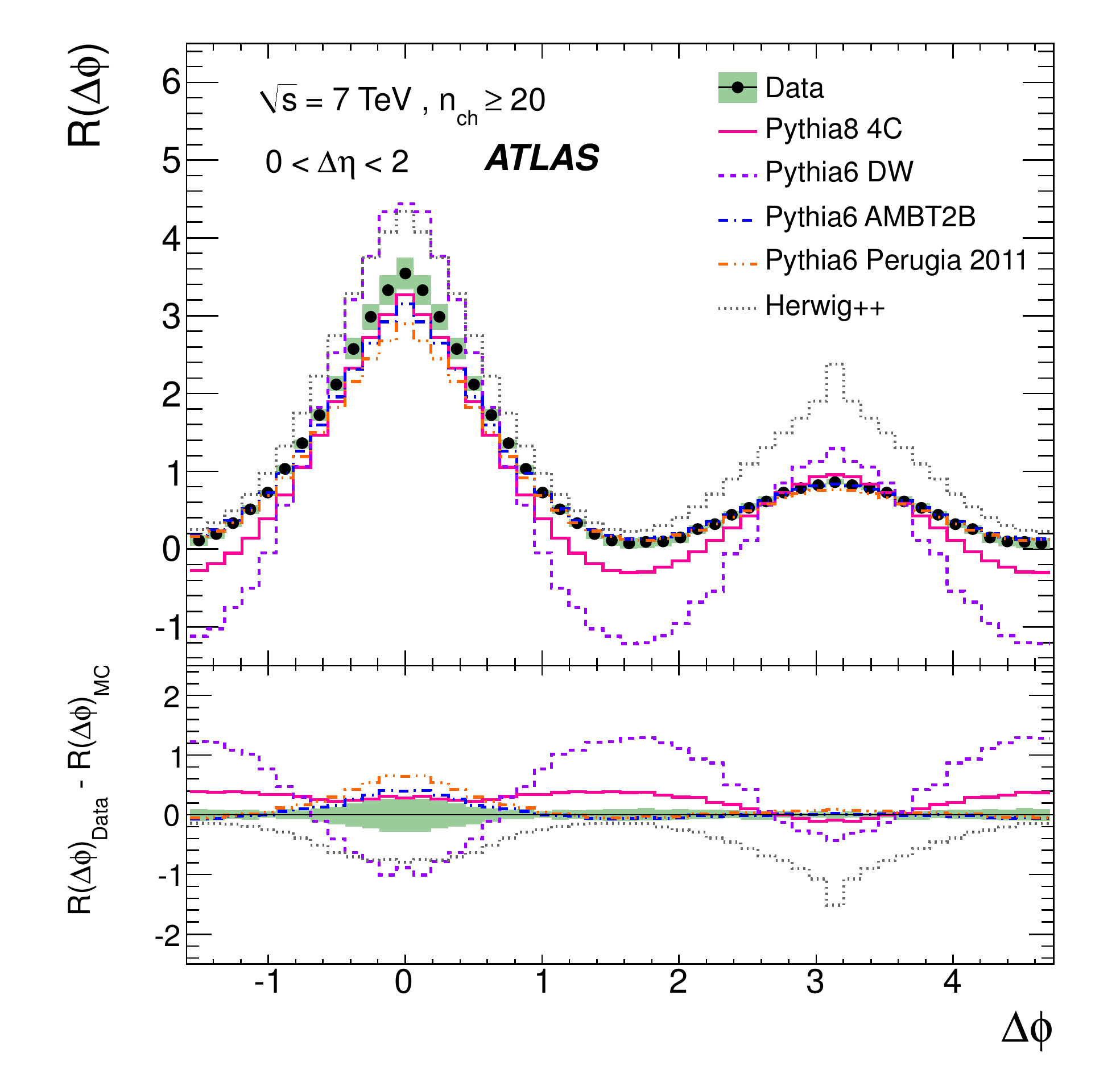}
\put(85,85){(c)}
\end{overpic}
\caption{Corrected $R(\Delta\phi$) two-particle correlation functions obtained by integrating $\Delta\eta$ between 0 and 2 for data and the different Monte Carlo tunes at (a) $\sqrt{s} =$ 900 GeV and 7 TeV for events with a charged-particle multiplicity (b) $n_{ch}\geq2$ and (c) $n_{ch}\geq20$.}
\label{fig:dphi02}
\end{center}
\end{figure}

The dependence of the correlation function on the azimuthal separation can similarly be studied by integrating the foreground and background over $\Delta\eta$ for two different ranges: (0:2) (short-range correlations) and (2:5) (long-range correlations).  The short-range correlation functions, shown in figure \ref{fig:dphi02}, contain two peaks.  The first peak is at $\Delta\phi=0$ and becomes more pronounced if either the collision energy or the particle multiplicity is increased.  The second peak is at $\Delta\phi=\pi$ and has a height that is approximately independent of both the collision energy and the multiplicity.  We note the general similarity in the shape of these functions to some distributions obtained in measurements of underlying event; in particular the trough in the correlation function around $\Delta\phi\simeq\pi/2$ is somewhat analogous to the ``transverse" region of ref.~\cite{Aad:2010fh}.

None of the Monte Carlo models provides a good description of the $\Delta\phi$ dependence of the short-range correlation function across the entire $\Delta\phi$ range. AMBT2B and Perugia~2011 tend to agree with data on the away side, but less so on the near side. DW, which has previously provided reasonable agreement with the transverse region of underlying event data, does not agree particularly well with the correlation function.  Tune 4C of \pythia~8 shows good agreement with data in some regions near the two peaks, but a clear disagreement in others, particularly the region around $\Delta\phi\simeq\pi/2$ at $\sqrt{s} =$ 7~TeV. 

\begin{figure}[t]
\begin{center}
\begin{overpic}[height=70mm]{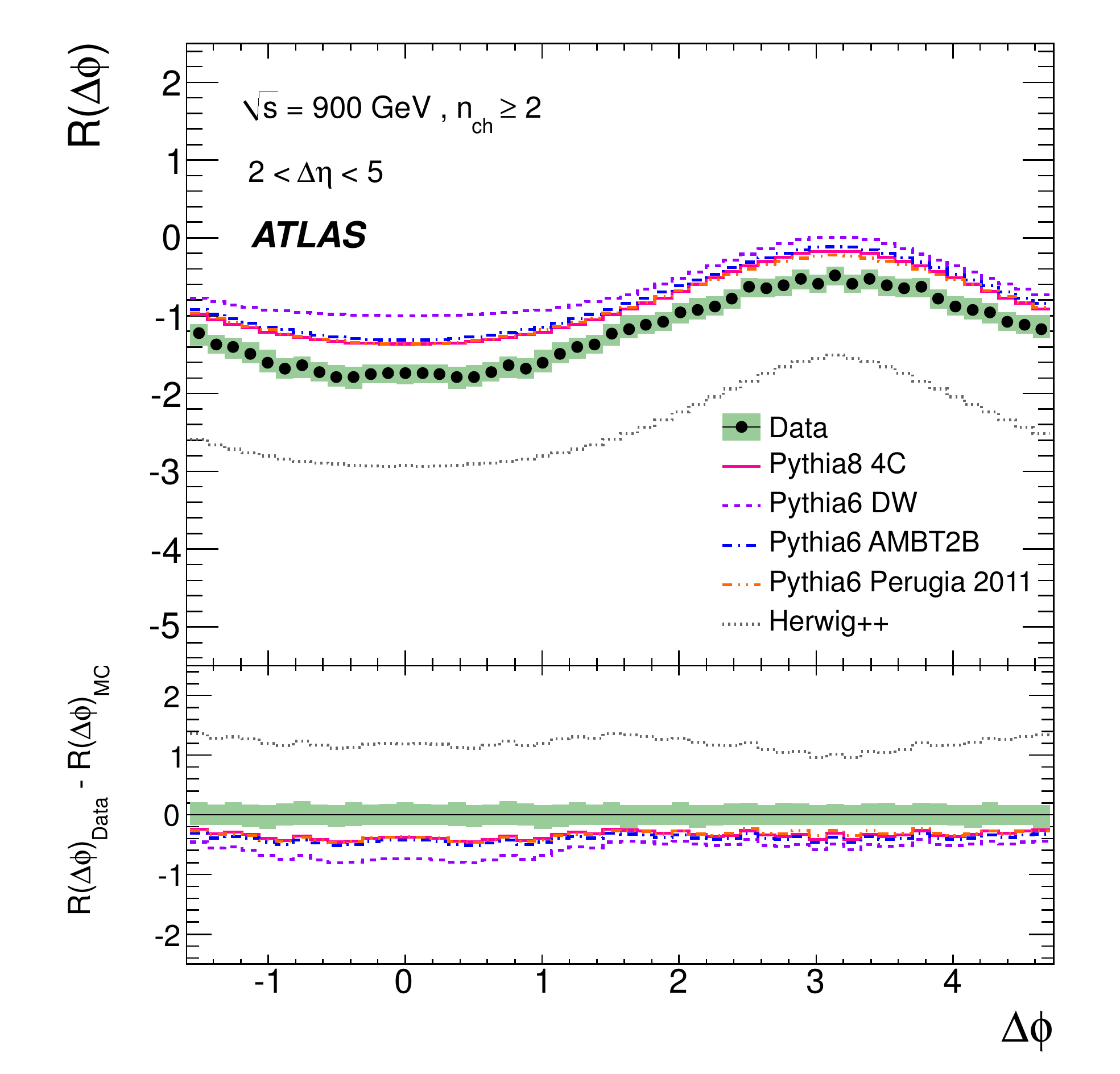}
\put(85,85){(a)}
\end{overpic} \\
\begin{overpic}[height=70mm]{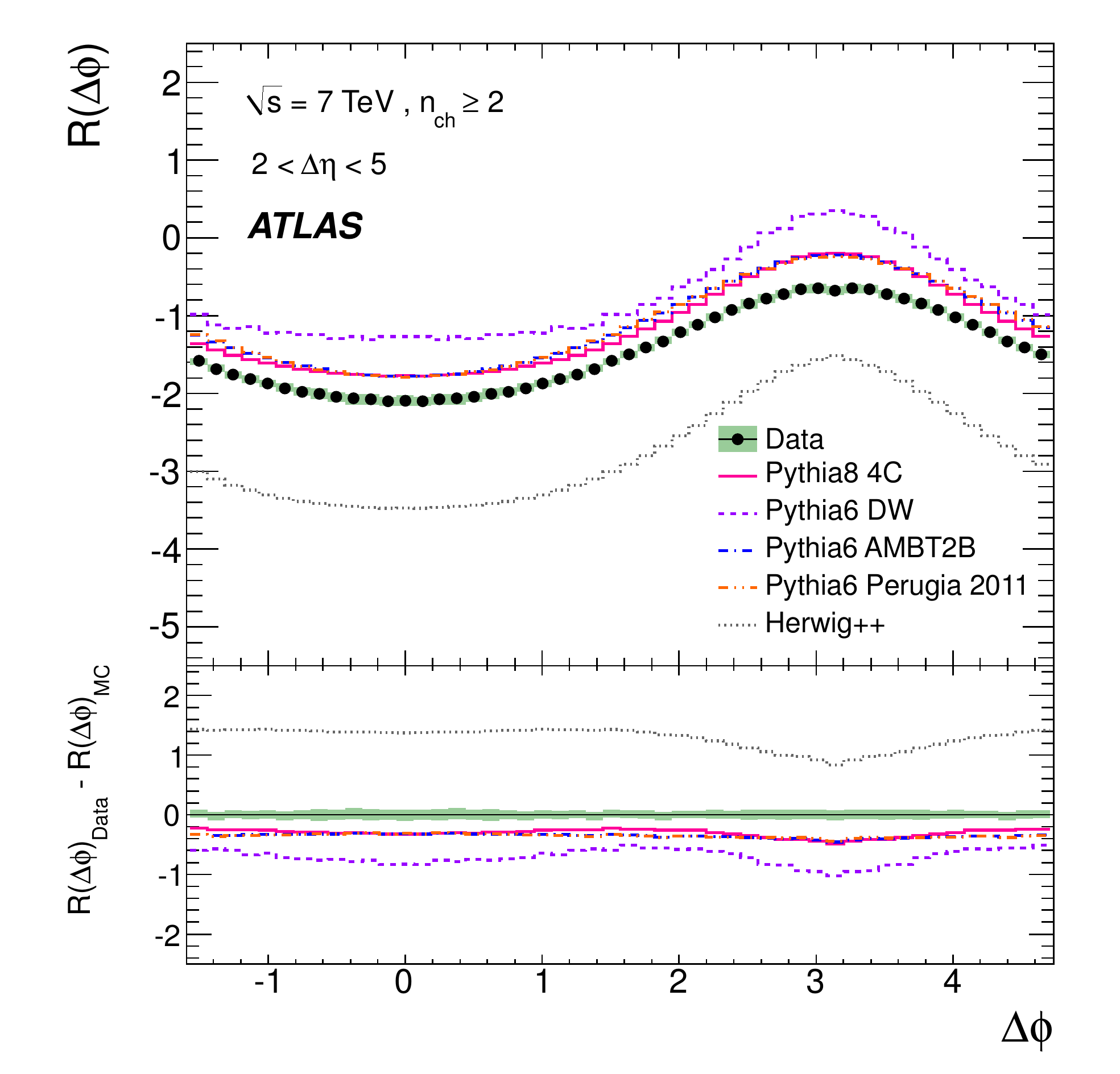}
\put(85,85){(b)}
\end{overpic}
\begin{overpic}[height=70mm]{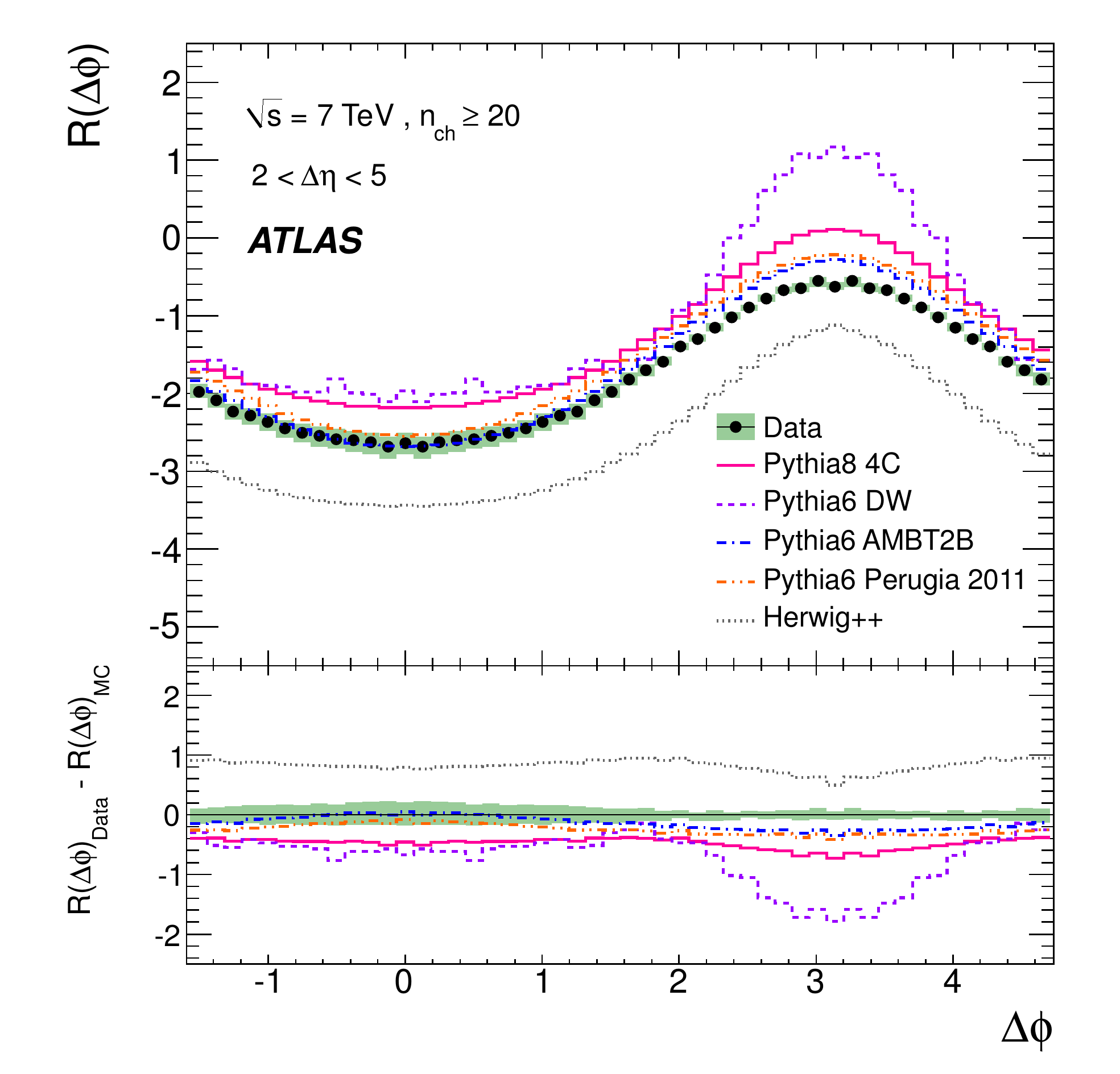}
\put(85,85){(c)}
\end{overpic}
\caption{Corrected $R(\Delta\phi$) two-particle correlation functions obtained by integrating the foreground and background distributions over $\Delta\eta$ between 2 and 5 for data and the different Monte Carlo tunes at (a) $\sqrt{s} =$ 900 GeV and 7 TeV for events with a charged-particle multiplicity (b) $n_{ch}\geq2$ and (c) $n_{ch}\geq20$.}
\label{fig:dphi25}
\end{center}
\end{figure}

Unlike the short-range correlation function, the $\Delta\phi$ dependence of the long-range correlation function does not exhibit a peak at $\Delta\phi=0$.  This is shown in figure \ref{fig:dphi25}, which reveals a trough at $\Delta\phi=0$ while the away-side peak is still present.  With the exception of \pythia~6 tune DW, the difference between the Monte Carlo models and the data is approximately flat in $\Delta\phi$.  While the models here describe the \emph{shape} of the correlation function, they do not describe the \emph{strength} of the correlations.  AMBT2B, Perugia~2011 and 4C are the best of the models considered here, although 4C diverges slightly more than the other two tunes when compared to the $n_{ch}\geq20$ sample of events. This $\Delta\phi$ projection was recently studied in high-multiplicity events by the CMS Collaboration and a ridge-like structure was observed at $\Delta\phi = 0$ in a specific kinematic range, different to the one studied here \cite{CMS:2010}. 

CMS has also studied the $\pT$-inclusive two-particle angular correlation function at $\sqrt{s} =$~900~GeV and 7 TeV. Their results \cite{CMS:2010} include the two-dimensional distributions, $R(\Delta\eta,\Delta\phi)$, and a single $R(\Delta\eta)$ projection where the correlation function has been integrated over $\Delta\phi$  between 0 and $\pi$. The phase-space used in their analysis is $\pT > 100$ MeV and $|\eta| < 2.4$, a difference of 0.1 units in pseudorapidity with respect to the analysis presented in this paper.  In comparing the $\Delta\eta$ dependence of the correlation function obtained by CMS with the results shown here, one should note a small and subtle difference in the definitions of the observables used in the two analyses. Whereas the analysis presented here measures the foreground, background and final correlation function inclusively for all event multiplicities within the event selection, the CMS analysis first determines the correlation function in separate bins of multiplicity.  The weighted average of the function in each bin is then computed to give an inclusive set of distributions. In order to see good agreement between these two observable definitions, one would need to correct the correlation function in each multiplicity bin when using the CMS definition \emph{and} correct the multiplicity selection requirements used to define the multiplicity-dependent correlation function. The analysis performed by CMS was reproduced using the data presented here, and the results obtained were found to be in good agreement with those published by the CMS Collaboration.

\section{Summary and conclusions}
\label{sec:summary}

The two-particle angular correlation functions in $\Delta\eta$ and $\Delta\phi$ have been measured using charged particles with $\pT >$ 100~MeV and $|\eta| <$ 2.5. Events were recorded by the ATLAS detector at the LHC from proton-proton collisions at $\sqrt{s} =$ 900 GeV and 7 TeV, using a minimum bias trigger. In total, integrated luminosities of approximately 7~$\mu\mathrm{b}^{-1}$ and 190~$\mu\mathrm{b}^{-1}$ have been analysed, respectively. At $\sqrt{s} =$ 7 TeV, two different phase-space regions defined by the charged particle multiplicity were investigated: $n_{ch}\geq2$ and $n_{ch}\geq20$. The structure of the correlation was explored in more detail by projecting the two-dimensional distributions into both $\Delta\eta$ and $\Delta\phi$. The results have been compared to Monte Carlo predictions, which show a similar complex structure but fail to reproduce the strength of the correlations seen in the data. Overall, the AMBT2B tune of \pythia\ 6 and the 4C tune of \pythia\ 8 are the models that provide the best description, while \herwig++ is the most discrepant in all of the distributions.  

While many of the models shown here reproduce the general features of the two-particle correlation function, none of them provide a good quantitative description of the strength of the correlations. In order to properly describe the correlations, the phenomenology of soft particle production needs further improvement.  Such improvement will, at a minimum, require the adjustment of existing models for diffraction and hadronisation.  However, it may be that the changes required to describe these data will extend beyond the re-tuning of existing models.

\acknowledgments

We thank CERN for the very successful operation of the LHC, as well as the
support staff from our institutions without whom ATLAS could not be
operated efficiently.

We acknowledge the support of ANPCyT, Argentina; YerPhI, Armenia; ARC,
Australia; BMWF, Austria; ANAS, Azerbaijan; SSTC, Belarus; CNPq and FAPESP,
Brazil; NSERC, NRC and CFI, Canada; CERN; CONICYT, Chile; CAS, MOST and NSFC,
China; COLCIENCIAS, Colombia; MSMT CR, MPO CR and VSC CR, Czech Republic;
DNRF, DNSRC and Lundbeck Foundation, Denmark; EPLANET and ERC, European Union;
IN2P3-CNRS, CEA-DSM/IRFU, France; GNAS, Georgia; BMBF, DFG, HGF, MPG and AvH
Foundation, Germany; GSRT, Greece; ISF, MINERVA, GIF, DIP and Benoziyo Center,
Israel; INFN, Italy; MEXT and JSPS, Japan; CNRST, Morocco; FOM and NWO,
Netherlands; RCN, Norway; MNiSW, Poland; GRICES and FCT, Portugal; MERYS
(MECTS), Romania; MES of Russia and ROSATOM, Russian Federation; JINR; MSTD,
Serbia; MSSR, Slovakia; ARRS and MVZT, Slovenia; DST/NRF, South Africa;
MICINN, Spain; SRC and Wallenberg Foundation, Sweden; SER, SNSF and Cantons of
Bern and Geneva, Switzerland; NSC, Taiwan; TAEK, Turkey; STFC, the Royal
Society and Leverhulme Trust, United Kingdom; DOE and NSF, United States of
America.

The crucial computing support from all WLCG partners is acknowledged
gratefully, in particular from CERN and the ATLAS Tier-1 facilities at
TRIUMF (Canada), NDGF (Denmark, Norway, Sweden), CC-IN2P3 (France),
KIT/GridKA (Germany), INFN-CNAF (Italy), NL-T1 (Netherlands), PIC (Spain),
ASGC (Taiwan), RAL (UK) and BNL (USA) and in the Tier-2 facilities
worldwide.

\bibliographystyle{JHEP}

\begin{thebibliography}{99}

\bibitem{Andersson:1987}
B. Andersson{ et al.}, {\it A model for low-$\pT$ hadronic reactions with generalizations to hadron-nucleus and nucleus-nucleus collisions}, \href{http://dx.doi.org/10.1016/0550-3213(87)90257-4}{{\em Nucl. Phys.} {\bf B281} (1987) 289}.

\bibitem{Morel:1974}
A. Morel and G. Plaut, {\it How do clusters look like in semi-inclusive cross sections?}, \href{http://dx.doi.org/10.1016/0550-3213(74)90597-5}{{\em Nucl. Phys.} {\bf B78} (1974) 541}.

\bibitem{Eggert:1975}
K. Eggert{ et al.}, {\it Angular Correlations Between the Charged Particles Produced in pp Collisions at ISR Energies}, \href{http://dx.doi.org/10.1016/0550-3213(75)90440-X}{{\em Nucl. Phys.} {\bf B86} (1975) 201}.

\bibitem{UA5:1988}
UA5 Collaboration, {\it Charged particle correlations in anti-p p collisions at c.m. energies of 200, 546 and 900 GeV}, \href{http://dx.doi.org/10.1007/BF01579906}{{\em Z. Phys.} {\bf C37} (1988) 191}.

\bibitem{PHOBOS:2007}
PHOBOS Collaboration, {\it Cluster properties from two-particle angular correlations in p+p collisions at $\sqrt{s} =$ 200 and 410 GeV}, \href{http://dx.doi.org/10.1103/PhysRevC.75.054913}{{\em Phys. Rev.} {\bf C75} (2007) 054913}.

\bibitem{ATLAS:2008}
ATLAS Collaboration, {\it The ATLAS Experiment at the CERN Large Hadron Collider}, \href{http://dx.doi.org/10.1088/1748-0221/3/08/S08003}{{\em JINST} {\bf 3} (2008) S08003}.

\bibitem{LHC:2008}
L. Evans and P. Bryant (eds.), {\it LHC Machine}, \href{http://dx.doi.org/10.1088/1748-0221/3/08/S08001}{{\em JINST} {\bf 3} (2008) S08001}.

\bibitem{Pythia6:2006}
T. Sj\"ostrand, S. Mrenna, and P. Skands, {\it PYTHIA 6.4 Physics and Manual}, \href{http://dx.doi.org/10.1088/1126-6708/2006/05/026}{{\em JHEP} {\bf 05} (2006) 026}.

\bibitem{Pythia8:2008}
T. Sj\"ostrand, S. Mrenna, and P. Skands, {\it Brief Introduction to Pythia 8.1}, \href{http://dx.doi.org/10.1016/j.cpc.2008.01.036}{{\em Comput. Phys. Comm.} {\bf 178} (2008) 852}.

\bibitem{Herwig:2008}
M. B\"ahr{ et al.}, {\it Herwig++ Physics and Manual}, \href{http://dx.doi.org/10.1140/epjc/s10052-008-0798-9}{{\em Eur. Phys. J.} {\bf C58} (2008) 639}.

\bibitem{HSoft:2009}
M. B\"ahr{ et al.}, {\it Soft Interactions in Herwig++}, \href{http://xxx.lanl.gov/abs/0905.4671}{{\tt arXiv:0905.4671 [hep-ph]}}.

\bibitem{AMBT:2011}
ATLAS Collaboration, {\it ATLAS Tunes of Pythia 6 and Pythia 8 for MC11}, {\scshape ATL-PHYS-PUB-2011-009}, \url{http://cdsweb.cern.ch/record/1363300}.

\bibitem{Per11:2011}
P. Skands, {\it Tuning Monte Carlo Generators: The Perugia Tunes}, \href{http://xxx.lanl.gov/abs/1005.3457}{{\tt arXiv:1005.3457 [hep-ph]}}.

\bibitem{DW:2002}
M. G. Albrow{ et al.}, {\it Tevatron-for-LHC Report of the QCD Working Group}, \href{http://xxx.lanl.gov/abs/hep-ph/0610012}{{\tt hep-ph/0610012}}.

\bibitem{P84C:2011}
R. Corke and T. Sj\"ostrand, {\it Interleaved parton showers and tuning prospects}, \href{http://dx.doi.org/10.1007/JHEP03(2011)032}{{\em JHEP} {\bf 03} (2011) 032}.

\bibitem{HTunes:2011}
\url{http://projects.hepforge.org/herwig/trac/wiki/MB_UE_tunes}.

\bibitem{H25:2011}
S. Gieseke{ et al.}, {\it Herwig++ 2.5 Release Note}, \href{http://xxx.lanl.gov/abs/1102.1672}{{\tt arXiv:1102.1672 [hep-ph]}}.

\bibitem{CTEQ6:2002}
J. Pumplin{ et al.}, {\it New generation of parton distributions with uncertainties from global QCD analysis}, \href{http://dx.doi.org/10.1088/1126-6708/2002/07/012}{{\em JHEP} {\bf 07} (2002) 012}.

\bibitem{MRTS:2008}
A. Sherstnev and R. S. Thorne, {\it Parton distributions for LO generators}, \href{http://dx.doi.org/10.1140/epjc/s10052-008-0610-x}{{\em Eur. Phys. J.} {\bf C55} (2008) 553}. 

\bibitem{Per0:2009}
P. Skands, {\it The Perugia Tunes}, \href{http://xxx.lanl.gov/abs/0905.3418}{{\tt arXiv:0905.3418 [hep-ph]}}.

\bibitem{MC09:2010}
ATLAS Collaboration, {\it ATLAS Monte Carlo Tunes for MC09}. {\scshape ATL-PHYS-PUB-2010-002}, \url{http://cdsweb.cern.ch/record/1247375}.

\bibitem{Phojet:1995}
R. Engel, {\it Photoproduction within the two component dual parton model: Amplitudes and cross-sections}, \href{http://dx.doi.org/10.1007/BF01496594}{{\em Z. Phys.} {\bf C66} (1995) 203}.

\bibitem{Geant4:2002}
GEANT4 Collaboration, {\it GEANT4: A simulation toolkit}, \href{http://dx.doi.org/10.1016/S0168-9002(03)01368-8}{{\em Nucl. Instr. Meth.} {\bf A506} (2003) 250}.

\bibitem{ATLASsim:2010}
ATLAS Collaboration, {\it The ATLAS Simulation Infrastructure}, \href{http://dx.doi.org/10.1140/epjc/s10052-010-1429-9}{{\em Eur. Phys. J} {\bf C70} (2010) 823}.

\bibitem{MB2:2011}
ATLAS Collaboration, {\it Charged-particle multiplicities in pp interactions measured with the ATLAS detector at the LHC}, \href{http://dx.doi.org/10.1088/1367-2630/13/5/053033}{{\em New J. Phys.} {\bf 13} (2011) 053033}.

\bibitem{Monk:2011pg}
J. Monk and C. Oropeza-Barrera, {\it The HBOM Method for Detector Corrections}, submitted to Nucl. Instr. Meth. (2011), \href{http://xxx.lanl.gov/abs/1111.4896}{{\tt arXiv:1111.4896 [hep-ex]}}.

\bibitem{Aad:2010fh}
ATLAS Collaboration, {\it Measurement of underlying event characteristics using charged particles in pp collisions at $\sqrt{s} = 900 GeV$ and 7 TeV with the ATLAS detector}, \href{http://dx.doi.org/10.1103/PhysRevD.83.112001}{{\em Phys. Rev.} {\bf D83} (2011) 112001}.

\bibitem{CMS:2010}
CMS Collaboration, {\it Observation of Long-Range, Near-Side Angular Correlations in Proton-Proton Collisions at the LHC}, \href{http://dx.doi.org/10.1007/JHEP09(2010)091}{{\em JHEP} {\bf 09} (2010) 091}.

\end{thebibliography}

\providecommand{\href}[2]{#2}\begingroup\raggedright
\endgroup

\clearpage 
% ATLAS Collaboration author list for 06-DEC-2011
% Data extracted on 16-Mar-2012 for paperid 167
%\documentclass[11pt]{article}
%\usepackage{a4wide}\begin{document}
\begin{flushleft}
{\Large The ATLAS Collaboration}

\bigskip

G.~Aad$^{\rm 48}$,
B.~Abbott$^{\rm 110}$,
J.~Abdallah$^{\rm 11}$,
A.A.~Abdelalim$^{\rm 49}$,
A.~Abdesselam$^{\rm 117}$,
O.~Abdinov$^{\rm 10}$,
B.~Abi$^{\rm 111}$,
M.~Abolins$^{\rm 87}$,
O.S.~AbouZeid$^{\rm 157}$,
H.~Abramowicz$^{\rm 152}$
% ...
\bigskip

% ...
$^{10}$ Institute of Physics, Azerbaijan Academy of Sciences, Baku, Azerbaijan\\
$^{11}$ Institut de F\'isica d'Altes Energies and Departament de F\'isica de la Universitat Aut\`onoma  de Barcelona and ICREA, Barcelona, Spain\\
% ...
$^{48}$ Fakult\"{a}t f\"{u}r Mathematik und Physik, Albert-Ludwigs-Universit\"{a}t, Freiburg i.Br., Germany\\
$^{49}$ Section de Physique, Universit\'e de Gen\`eve, Geneva, Switzerland\\
% ...
$^{87}$ Department of Physics and Astronomy, Michigan State University, East Lansing MI, United States of America\\
% ...
$^{110}$ Homer L. Dodge Department of Physics and Astronomy, University of Oklahoma, Norman OK, United States of America\\
$^{111}$ Department of Physics, Oklahoma State University, Stillwater OK, United States of America\\
% ...
$^{117}$ Department of Physics, Oxford University, Oxford, United Kingdom\\
% ...
$^{152}$ Raymond and Beverly Sackler School of Physics and Astronomy, Tel Aviv University, Tel Aviv, Israel\\
% ...
$^{157}$ Department of Physics, University of Toronto, Toronto ON, Canada\\
% ...
$^{b}$ Also at Faculdade de Ciencias and CFNUL, Universidade de Lisboa, Lisboa, Portugal\\
% ...
$^{s}$ Also at School of Physics, Shandong University, Shandong, China\\
% ...
\end{flushleft}

%\end{document} 

\end{document}